**Spatiotemporally-resolved differential dynamic microscopy reveals strain-induced deformations in heterogeneous soft materials**

Aysan Razzaghi[1], Mehrzad Sasanpour[1], Jasmine Ren[1], Caidric Gupit[2,3,4], Matthew E. Helgeson[3], Megan T. Valentine[2], Rae M. Robertson-Anderson[1]*, Ryan McGorty[1]*

*corresponding authors, contributed equally: rmcgorty@sandiego.edu, randerson@sandiego.edu

[1] Department of Physics and Biophysics, University of San Diego

[2] Department of Mechanical Engineering, University of California, Santa Barbara

[3] Department of Chemical Engineering, University of California, Santa Barbara

[4] Present affiliation: Institute of Chemistry, University of the Philippines Diliman, Quezon City, Philippines

## Abstract

Understanding the microscopic behavior of materials in response to mechanical perturbations is a central challenge in soft matter, yet existing methods, which typically rely on microscopy-based imaging, often struggle with spatial heterogeneity and low signal-to-noise that are characteristic of many soft and biological materials. Additionally, many existing rheo-optical methods focus on the response to macro-scale forces rather than localized perturbations despite the importance of local stresses in biological and industrial applications of soft materials. Here, we present STroboscopically Averaged, time-Resolved Differential Dynamic Microscopy (starDDM) to investigate a sample's deformation across time and distance due to a cyclically applied local force. As proof-of-principle, we use this model-independent and tracer-free approach to map the motions of active cytoskeleton composites imposed by the movement of an optically trapped probe particle embedded within the complex network. We reveal how motor-driven restructuring and varying strain rates tune the magnitude and propagation of the response, which we further show decays with distance from the perturbation. This analytical platform fills a critical gap in micro-rheology, offering a versatile tool for probing the non-equilibrium dynamics of soft materials ranging from biopolymer networks to 3D-printed materials and soft robotics.

## Introduction

Understanding how soft materials respond to deformations impacts fields from biophysics to materials science. Identifying the relationships between rheological properties, sample structure, and microscopic dynamics is a central concern toward this aim. To reveal these relationships, researchers have devised many innovative rheo-optical methods that probe local sample microstructure, rheology, velocity fields, composition and microscale dynamics.[1-18] These methods typically combine mechanical perturbation of the sample with simultaneous measurement of structure using microscopic imaging, scattering methods, or a hybrid approach. The use of microscopy in rheo-optical methods has the advantages of spatially-resolved dynamics and compatibility with various contrast-generating methods (e.g., fluorescence or phase contrast). Scattering techniques, meanwhile, provide ensemble-averaged dynamics over a wide range of time and length scales that can be straightforwardly compared to theories on the condition that the length scales of the constituent material elements are well separated from the length scales of their concentration fluctuations. In recent years, new optical methods, combining features of both microscopy and scattering, have been employed in rheo-optical setups. These optical techniques have been categorized as near-field techniques[19] and include differential dynamic microscopy (DDM) which uses real-space optical microscopy videos to analyze density fluctuations in Fourier space.[20-22]

Often, these rheo-optical methods are used in situations where the spatiotemporal sample deformation pattern is relatively simple, such as with steady shear or sinusoidal oscillations applied homogeneously at the macroscopic scale.[1,13,23,24] For example, “Echo-DDM” and “Flow-DDM” use DDM strategies to characterize the shear-induced dynamics in materials under bulk oscillatory[18] or steady flow[25], leveraging analysis frameworks that decouple the affine and non-affine motion[26] of a material under shear. These techniques specifically aim to separate the component of the motion that aligns with the strain (affine) to instead investigate the (non-affine) fluctuations of the material constituents in the reference frame of the strain. As such, their applicability is limited to materials that have ample thermal or diffusive fluctuations on the timescale of the strains that DDM can resolve, which is often not the case in materials that are primarily elastic. They also rely on assumptions about the flow fields induced by the imposed deformation, which are non-trivial in cases in which the perturbation is localized. While previous work has used particle tracking based methods to quantify responses to local perturbations,[27,28] such methods often require user-defined analysis criteria, making them time-consuming and difficult to automate; and existing extensions of DDM cannot be directly used to determine how the material deforms over time and as a function of distance from a microscopic perturbation.

Here, we take an orthogonal approach to these existing methods. We focus specifically on the strain-induced motion of the material due to localized perturbations that lack an analytically tractable description of the resulting flow field. With this approach, we circumvent the need for measurable random or non-affine fluctuations and can extract meaningful information about the response of materials with a high degree of spatial heterogeneity, or under conditions with low signal-to-noise ratios — features that are common in soft, active, and biological materials. The STroboscopically-Averaged time-Resolved DDM (starDDM) method we introduce here focuses not on the thermally-driven correlations of density fluctuations, as is done in typical DDM applications, but on their time-varying spatial anisotropy due to the material's mechanical response to a locally imposed strain. By examining the motion of the material in regions of interest (ROIs) that are fixed relative to the time-varying strain pattern, starDDM can resolve information regarding the local strain field in a material. Importantly, unlike other DDM microrheology approaches,[29-31] we do not aim to quantify conventional rheological metrics, such as viscoelastic moduli, which are inherently difficult to interpret for heterogeneous non-continuum materials. Rather, starDDM provides a phenomenological framework for characterizing the dynamic response of soft materials to perturbations in cases where the strain field is complex or not known and the material cannot be described as a continuum elastic or viscoelastic material on the scale of the perturbation.

A key benefit of using a digital Fourier microscopy method like DDM is that it provides advantages of scattering approaches using an optical microscope as a platform, which we leverage here through integration with optical tweezers (OT) that provide precisely programmed local mechanical perturbations. Existing microscopy image analysis techniques, such as single particle tracking or particle image velocimetry, could in principle characterize deformation patterns within the sample; however, they often rely on tracking foreign tracer particles or require prohibitively high image contrast. By comparison, DDM offers distinct advantages: it is compatible with multiple microscopy modalities (e.g., brightfield, fluorescence, phase), does not require tracer probes, and performs well under low signal-to-noise[32,33] conditions where particle tracking methods often fail, making it particularly suitable for characterizing dynamics in systems in which features are not clearly optically resolved.

We previously showed that combining DDM with OT-based material forcing could be used to relate the local force response to the macromolecular dynamics and strain propagation in fluids of biopolymers.[34,35] In these studies, we showed that the alignment factor ($AF$) – characterizing the degree of anisotropy in the two-dimensional dynamic structure function aligned with the strain direction – could describe how the local affine deformation (in the direction of the strain) within the material decays with distance from a locally imposed force. In these previous studies, we examined dense solutions of labeled DNA molecules which were spatially homogeneous, uniformly filling each ROI with high signal-to-noise and ample statistics. The molecules also displayed measurable thermal fluctuations that could be used to extract dynamics using standard DDM approaches.

Here, we introduce starDDM to extend this framework to spatiotemporally heterogeneous and more solid-like materials that do not have these optimal characteristics and show that we can determine how local mechanical perturbations deform soft materials across space and time. Such analyses can provide important insight for the use of such materials in both biological and industrial applications. We choose as our proof-of-concept experimental system composites of actin filaments and microtubules that have been restructured by tetrameric kinesin motors capable of crosslinking and sliding microtubule filaments. Such composites exhibit spatiotemporal heterogeneity in both structure and mechanics, and minimal optically-resolvable thermal fluctuations due to the high degree of elasticity relative to dissipative response features. Prior studies have established their complex structure-mechanics relationships including non-monotonic dependence of composite elasticity on kinesin concentration due to emerging structural heterogeneity and network demixing.[36] However, due to the limitations described above, existing methods could not quantitatively assess how deformations due to local stresses propagated through the material, which is highly relevant to not only understanding foundational cellular and tissue mechanics but to applications and industrial processes as diverse as bioprinting and soft robotics.

## Results and Discussion

We aim to develop, validate and optimize starDDM to extract information about spatiotemporally varying deformation fields from noisy microscopy data of heterogeneous materials. We demonstrate the method through experiments where we impose a precise temporally patterned disturbance to cytoskeleton composites by cyclically displacing an optically trapped microsphere (radius $r_p$ = 2.25 μm) through a displacement of $x = 20$ μm (Fig. 1(a)). The initial stress-free state corresponds to the trapped particle positioned at ($x$ , $y$) = (−10 μm, 0 μm). The trapped particle is then displaced along the $x$-axis to position (+10 μm, 0 μm) and back at constant speed. We simultaneously acquire time-series of images of fluorescently labeled filaments comprising the network in a field-of-view (FOV) centered on the optical trap and extending vertically to $y = 56$ μm. The applied force visibly deforms the network in the region surrounding the microsphere, and the magnitude of deformation appears to decrease with distance from the moving microsphere (SI Movies S1-S10). Strong spatial heterogeneity in composite structure and regions of low signal-to-noise are also apparent throughout the FOV and from trial to trial and across replicates, making it difficult to resolve possible trends through simple inspection. Tracking probes embedded within the network is challenging due to the presence of large pores through which inert probes can diffuse as they sample the interstitial fluid permeating the meshwork rather than the network itself. Adhesive particles could in principle track the network motion, but there is a risk that they would nucleate filament aggregates or modify the structure. PIV or optical flow measurements have been shown to provide useful information about overall flow, but fail to isolate the large-scale motions arising from long-range stress transmission that we demonstrate can be selectively accessed with starDDM.

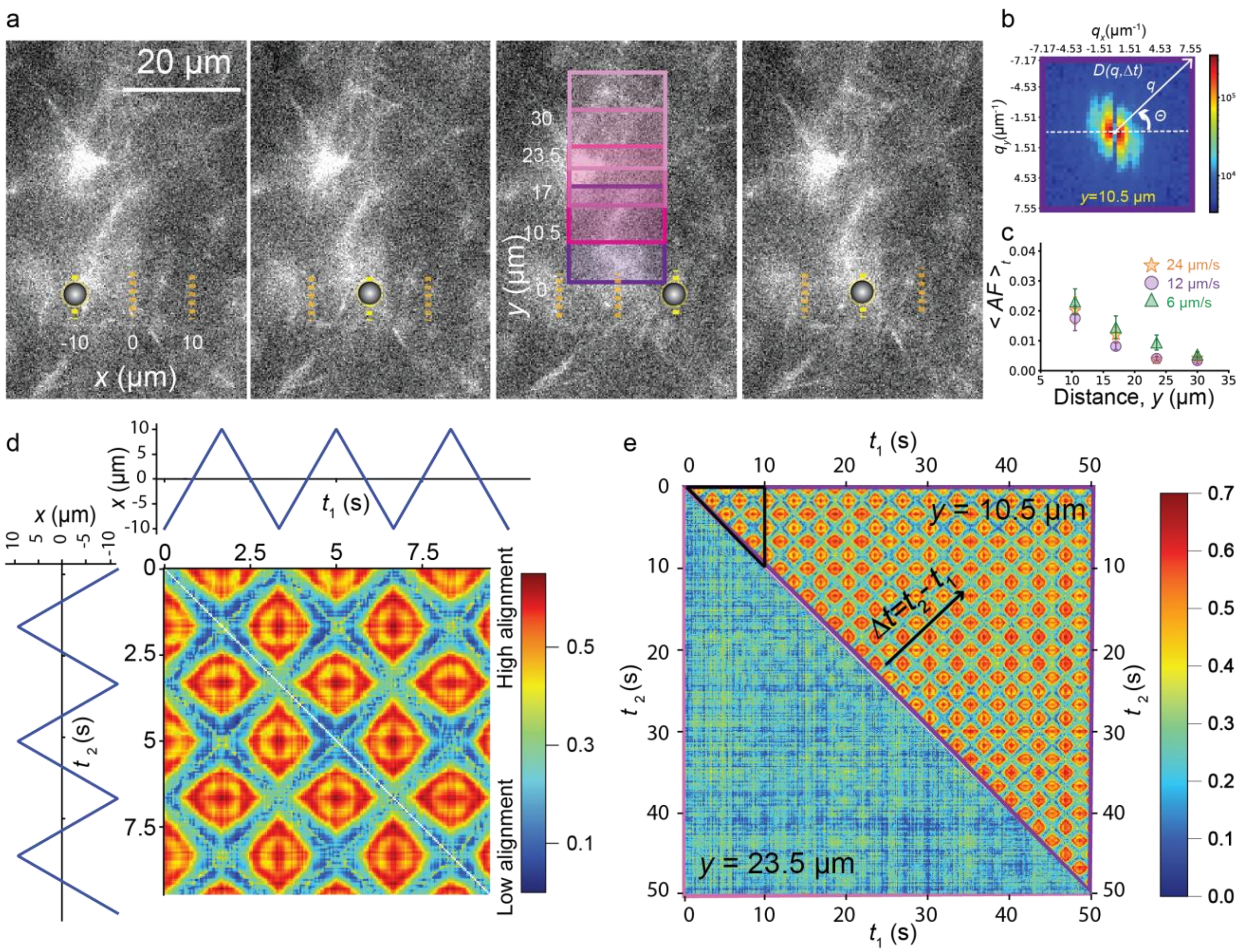

**Figure 1. Time-resolved DDM maps the deformation field of a cytoskeleton composite subjected to local perturbations**. **(a)** An embedded spherical probe is optically swept back and forth through a displacement 20 μm at constant speed along the horizontal ($x$) direction through a cytoskeleton composite while videos of the labeled filaments (actin and microtubules) are simultaneously captured with epifluorescence microscopy. The four panels show sample video frames, separated by equal time intervals in one complete cycle ($x = -10$, 0, +10, 0 μm). The initial stress-free state corresponds to ($x$ , $y$) position (−10, 0 μm). To determine the dependence of filament motion on vertical distance from the cyclically-swept optical trap, we perform time-

resolved DDM on 4 ROIs of area 16×16 μm$^2$ and centered horizontally at $x = 0$ and vertically at $y = 10.5$ (purple boxes), 17 (magenta), 23.5 (pink), 30 (light pink) μm, relative to the probe center at $y = 0$. This places the ROI window at the midpoint of particle motion per sweep. **(b)** Performing DDM on an ROI at $y = 10.5$ μm yields a 2D image structure function $D(\boldsymbol{q}, \Delta t)$ displayed as a heat map with higher values shown as warmer colors according to the color scale shown. Values of $D(\boldsymbol{q}, \Delta t)$ have been averaged over lag times of $\Delta t = 0.05\text{-}0.85\ s$. **(c)** To determine the extent to which the network motion aligns with the strain, we compute the alignment factor $\langle AF \rangle_t$ of $D(\boldsymbol{q}, \Delta t)$ relative to the strain path, as defined in Eqn. 3. Data shown is $\langle AF \rangle_t$ versus $y$ for strain speeds of $u$ (μm/s) = 6 (green triangles), 12 (purple circles) and 24 (orange stars) for each sampled ROI, measured for $q_r$ = 1.5 μm$^{-1}$. **(d)** The time-resolved alignment factor, $AF(q_r, t_1, t_2)$, measured for $q_r = 1.5$ μm$^{-1}$ and $y = 10.5$ μm is shown for the first 10 s of a video in which $u$ = 12 μm/s. The probe position at each $t_1$ and $t_2$ is shown along the horizontal and vertical axes. The top-left to bottom-right diagonal represents lag time $\Delta t = 0$ (i.e., $t_1 = t_2$) and increasing lag times run perpendicular to this diagonal. **(e)** $AF(q_r, t_1, t_2)$ is shown for an entire 50 s video that comprises 14 complete cycles for ROIs at $y = 10.5$ μm (top-right half) and $y = 23.5$ μm (bottom-left half). Data shown in (d) is in the top-left boxed-in region in (e).

We use spatially-resolved DDM to quantify the degree and spatial extent to which our composites deform in response to the cyclic strain. Specifically, we divide each FOV into overlapping regions of interest (ROIs) of 16 × 16 μm$^2$ (or 128 × 128 pixels) that are fixed in the experimental frame, centered at the midpoint of the path of the cyclically swept optical trap and spaced vertically above the strain path at 6.5 μm center-to-center intervals, indicated by colored squares in Fig. 1(a). Note that we chose ROIs that do not move with the optically trapped microsphere to explicitly capture the deformation induced by the moving probe. The motion we measure is a direct read-out of the strain-induced deformation of the material. This is distinct from approaches such as Echo-DDM[18,37] that aim to decouple the reversible deformations of the material and the shear-induced diffusive motion. A potential disadvantage of this approach is that the separation vector between the probe particle and ROI center varies throughout the cyclic sweep. However, as we show below, through judicious choice of cycle phase and lag time we are able to isolate the long length scale dynamics. In general, the network response to applied loading may vary with different ROI placements, especially for highly heterogenous materials, such as the cytoskeletal networks examined here.

For each ROI, we then perform DDM to reveal how intensity fluctuations over a range of length scales change in time. In detail, we compute the conventional image structure function or DDM matrix, $D(\mathbf{q}, \Delta t)$, defined as:

$$D(\mathbf{q}, \Delta t) = \langle \left| \tilde{I}(\mathbf{q}, t + \Delta t) - \tilde{I}(\mathbf{q}, t) \right|^2 \rangle_t, \qquad (1)$$

where $\tilde{I}(\mathbf{q}, t)$ is the spatial Fourier transform of the image, which is a function of time $t$, wave vector $\mathbf{q} = (q_x, q_y)$, and lag time $\Delta t$. We emphasize the vectorial nature of $\mathbf{q}$, which allows DDM to analyze anisotropic structure and dynamics.[18,25,26,35,37-42] The $\langle \cdots \rangle_t$ denotes an average over all time, appropriate when the dynamics are stationary, as is often the case with DDM analysis, though less appropriate in our experiments where the oscillatory deformation pattern imposes time-dependent dynamics. As we will describe later, we address this by performing time-resolved DDM analysis.

In the stationary case, $D(\mathbf{q}, \Delta t)$ can be modeled by the function

$$D(\mathbf{q}, \Delta t) = A(\mathbf{q})\{1 - f(\mathbf{q}, \Delta t)\} + B(\mathbf{q}), \tag{2}$$

where $A(\mathbf{q})$ is a static amplitude term that depends on the time-averaged structure of the objects in the image as well as on the transfer function of the optical system, $B(\mathbf{q})$ is a background term that depends on the uncorrelated background noise, and $f(\mathbf{q}, \Delta t)$ is the intermediate scattering function (ISF) that describes the dynamics of the material. Often in DDM analysis, this intermediate scattering function is fit to one of a variety of models analogous to those used to interpret dynamic correlation-based scattering measurements. For example, in the case of purely diffusive dynamics, one can use a model described by $f(q_r, \Delta t) = \exp(-Dq_r^2\Delta t)$ where $D$ is the diffusion coefficient and $q_r$ is the wave vector magnitude. Alternatively, an appropriate model for ballistic motion with random direction is $f(q_r, \Delta t) = \exp(-vq_r\Delta t)$ where $v$ is a characteristic velocity magnitude.

In our system, thermal fluctuations are not visibly present over the duration of the 50 second videos (SI Movies S1-S10). Rather, motion of the constituents of the material (i.e., actin filaments and microtubules) is primarily dominated by the oscillatory strain imposed by the motion of the optical tweezers. Consistent with this observation, the measured ISF exhibits periodic peaks that reflect the periodic motion of the trapped particle, and a pronounced anisotropy, with decay occurring more rapidly along the direction parallel to the probe motion rather than perpendicular to it (Fig. S1, SI Section S2).

While it may be possible to directly interpret the ISF in such cases, this approach requires a microstructurally-informed model, sufficient signal-to-noise to ensure accurate fitting, and the ability to estimate the parameters $A(\mathbf{q})$ and $B(\mathbf{q})$ or to include them as free parameters while fitting $D(\mathbf{q}, \Delta t)$. The strain field anisotropy may further require performing this procedure separately for $D(q_x, \Delta t)$ and $D(q_y, \Delta t)$ using small sectors of $\mathbf{q}$ values centered along the $q_x$ and $q_y$ axes in $\mathbf{q}$-space. We instead take a simpler, more data-driven approach, which requires no assumptions about the functional form of $f(\mathbf{q}, \Delta t)$ or $D(\mathbf{q}, \Delta t)$, thereby circumventing these limitations. Moreover, we aim to spatially and temporally resolve dynamics by executing the analysis on small ROIs within the captured field-of-view and over relatively short regions of time, where model fitting of the resulting $D(\mathbf{q}, \Delta t)$ may be challenging due to sampling noise.

Our approach focuses on evaluating the anisotropy of $D(\mathbf{q}, \Delta t)$, which we relate to the anisotropy of the material motion within the ROI as a result of the directional probe movement. In practice, we perform a weighted azimuthal average, which we refer to as the alignment factor, $\langle AF(q_r, \Delta t)\rangle_t$, consistent with prior work,[35,43,44] defined as

$$\langle AF\rangle_t = \int_0^{2\pi} D(q_r, \Delta t, \theta)\cos(2\theta)\ d\theta \Bigg/ \int_0^{2\pi} D(q_r, \Delta t, \theta)\ d\theta, \tag{3}$$

where $\theta$ defines an angle on the $(q_x, q_y)$ plane measured from the $x$-axis, as seen in Fig. 1(b). The $\langle\cdots\rangle_t$ brackets signify that this alignment factor is calculated on a time-averaged quantity, $D(\mathbf{q}, \Delta t)$.

When the resulting $\langle AF\rangle_t$ is calculated for ROIs at increasing distance from the path of the cyclically applied force, we find that $\langle AF\rangle_t$ decreases, in agreement with our qualitative observations that the visible deformation of the material decreases with distance from the probe (Fig. 1(c)). However, the use of the *time-averaged* $D(\mathbf{q}, \Delta t)$ likely diminishes our ability to accurately assess the anisotropic motion of the material. Given the oscillatory nature of the imposed probe motion, for any lag time, $\Delta t$, there will be instances within each oscillation cycle when the displacement of filaments within some ROI for that lag time is minimal and other times when, for the same lag time, the displacement is greater. That is, the degree to which the material within any fixed ROI deforms over some time interval, $\Delta t$, depends on *when* that time interval occurs relative to a strain cycle. Thus, by resolving the *time dependence* of the alignment factor rather than averaging over time, as is done in standard DDM implementations, we hypothesize that we can substantially increase the signal-to-noise of $AF$ *and* better resolve the time course of material dynamics during the strain cycle.

To achieve this, we introduce a time-dependent DDM matrix, $D(\mathbf{q}, t_1, t_2) = \left|\tilde{I}(\mathbf{q}, t_2) - \tilde{I}(\mathbf{q}, t_1)\right|^2$, from which we extract a time dependent alignment factor, $AF(q_r, t_1, t_2)$, which resembles a two-time correlation function[45,46] and conveys similar information about the time-dependence of the dynamics. $AF(q_r, t_1, t_2)$ is sensitive to asymmetries in $D(\mathbf{q}, t_1, t_2)$ that could arise from anisotropic dynamics, such as filaments moving in the direction of the applied force, or anisotropic structures such as filaments that preferentially align parallel or perpendicular to the strain field, or a combination of the two. To estimate the extent to which anisotropy in the structure dominates $AF$ in our experiments, we compute the image power spectrum, $\left|\tilde{I}(\mathbf{q}, t)^2\right|$, which is expected to be proportional to $A(\mathbf{q})$. We find that the anisotropy of $\left|\tilde{I}(\mathbf{q}, t)^2\right|$ is generally less than that of $D(\mathbf{q}, t_1, t_2)$ and does not vary with the strain cycle in the way that the anisotropy of $D(\mathbf{q}, t_1, t_2)$ does (Fig. S2). Therefore, for the subsequent analysis, we assume that the periodic dynamics observed in $AF$ are dominated

by the anisotropic straining motion of the material rather than by orientational anisotropy of the individual filaments, although we anticipate that effects from both anisotropic motion and anisotropic structure are present due to the high aspect ratio and relatively long contour lengths of the fluorescent filaments.

To more quantitatively link $AF$ to the motion of the material, we make some simplifying assumptions and consider some representative classes of materials (see details provided in SI Section 3). Assuming, as discussed above, that $A(\mathbf{q})$ is isotropic and that the background term, $B(\mathbf{q})$, is negligible then we can write the time-dependent alignment factor as:

$$AF(q_r, t_1, t_2) = -\frac{\frac{1}{2\pi}\int_0^{2\pi} f\,(\mathbf{q}, t_1, t_2)\cos(2\theta)\,d\theta}{1 - \frac{1}{2\pi}\int_0^{2\pi} f\,(\mathbf{q}, t_1, t_2)\,d\theta}. \quad (4)$$

Here, $f_a(\mathbf{q}, t_1, t_2)$ describes the contribution to the two-time intermediate scattering function due to advective motion in the material in response to the imposed motion of the probe. We assume that the contributions of non-affine or time-fluctuating motions to the intermediate scattering function are negligible. For purely advective motion of an incompressible material with a time- and space-dependent velocity field $\mathbf{v}(\mathbf{r}, t)$, $f_a(\mathbf{q}, t_1, t_2)$ is given by

$$f_a(\mathbf{q}, t_1, t_2) = \frac{1}{V}\int \exp\left[i\mathbf{q}\cdot\int_{t_1}^{t_2}\mathbf{v}\,(\mathbf{r}(t'|\mathbf{r}'), t')\,dt'\right]d^3\mathbf{r}'. \quad (5)$$

Here, $V$ is the ROI volume and $\mathbf{r}(t'|\mathbf{r}')$ is the Lagrangian path function describing the time-dependent location of a material element initially at $\mathbf{r}'$ subjected to the global velocity field $\mathbf{v}(\mathbf{r}, t)$. For the case of a probe moving in a material with a homogeneous and time-stationary viscoelastic response, $AF(q_r, t_1, t_2)$ can be generally related to measures of the local strain anisotropy within the material through

$$AF(q, t_1, t_2|\boldsymbol{\rho}) = \left(\frac{\Delta\delta_{\parallel}(\boldsymbol{\rho}, t_1, t_2)^2 - \Delta\delta_{\perp}(\boldsymbol{\rho}, t_1, t_2)^2}{\Delta\delta_{\parallel}(\boldsymbol{\rho}, t_1, t_2)^2 + \Delta\delta_{\perp}(\boldsymbol{\rho}, t_1, t_2)^2}\right)\frac{J_2\left(q_r\sqrt{\Delta\delta_{\parallel}(\boldsymbol{\rho}, t_1, t_2)^2 + \Delta\delta_{\perp}(\boldsymbol{\rho}, t_1, t_2)^2}\right)}{1 + J_0\left(q_r\sqrt{\Delta\delta_{\parallel}(\boldsymbol{\rho}, t_1, t_2)^2 + \Delta\delta_{\perp}(\boldsymbol{\rho}, t_1, t_2)^2}\right)}. \quad (6)$$

where $J_i(\ldots)$ is the $i^{\text{th}}$ order Bessel function of the first kind (see SI section 3). Here, $\Delta\delta_{\parallel}(\boldsymbol{\rho}, t_1, t_2)$ and $\Delta\delta_{\perp}(\boldsymbol{\rho}, t_1, t_2)$ are the difference in the parallel and perpendicular components (relative to the probe motion), respectively, of the difference in strain displacement field in an ROI separated by a vector $\boldsymbol{\rho}(t)$ from the time-dependent probe position. Thus, $AF(q_r, t_1, t_2)$ for a fixed ROI will contain information regarding the anisotropy in the local strain field in the vicinity of the probe particle as the probe is swept across the ROI.

The structures of Eqs. (5,6) show that $f_a(\mathbf{q}, t_1, t_2)$ and $AF(q_r, t_1, t_2)$ will generally have a complex dependence on time for a probe moving in a general viscoelastic material, the details of which are beyond the current scope. However, for the prototypical case of a probe moving with constant velocity $\boldsymbol{u}$ in a Hookean elastic solid, Eq. (5) can be simplified to the point of analytical solutions for $f_a(\mathbf{q}, t_1, t_2)$ and $AF(q_r, t_1, t_2)$ (see SI Section 3 for details). In this limit, the magnitude of $AF$ is expected to depend on the force, $F_0(t)$, needed to maintain the probe particle at its speed of $u$, the shear modulus $\mu$, and the Poisson ratio $\nu$ as:

$$\ln AF \propto \frac{F_0(t)\, u}{16\pi\mu(1-\nu)}. \qquad (7)$$

The actin filaments and microtubule networks considered in this work are expected to exhibit more complex viscoelastic dynamics and spatial heterogeneity, which will require further theoretical analysis. Nevertheless, Eqs. (6) and (7) provide some initial expectations for the analysis and interpretation of the active microrheology experiments to follow. First, we expect $AF$ to encode information regarding the structure of the local strain field anisotropy surrounding the driven probe particle. Specifically, for the case of cyclic probe motion and a fixed ROI in the experimental frame, we anticipate that $AF(q_r, t_1, t_2)$ will exhibit a cyclical time-symmetry. Furthermore, for an elastic-dominant material response, the magnitude of $AF$ is expected to scale with the probe velocity, and inversely with the distance from the probe in a way that will be sensitive to the local viscoelastic properties of the material. These expectations will be assessed in the analysis to follow.

The experimentally determined time dependence of the anisotropy in $D(\mathbf{q}, t_1, t_2)$ is clearly shown in Fig. 1(d), which shows $AF(q_r, t_1, t_2)$ for three strain oscillation cycles at a probe speed of $u = 12$ μm/s, evaluated for the ROI closest to the probe at its midpoint and at $q_r =$ 1.5 μm$^{-1}$, a $q_r$-value towards the lower end of our entire $q_r$-range of 0.38 to 23.8 μm$^{-1}$ where the signal-to-noise in the DDM matrix is higher. Here, there is a very clear periodicity commensurate with the period $T = 4x_{max}/u =$ 3.33 s of the strain. Peaks in $AF$ (red regions) occur when the lag time, $\Delta t$, is equal to half the period, in this case $T/2 = 1.67$ s, or integer multiples thereof, and when the intensity difference is calculated using the microsphere's position at the two extremal positions: $t_1$ corresponding to the time at which the particle is positioned at $-x_{max}$ and at $t_2$ corresponding to the time at which the particle is positioned at $+x_{max}$, or vice versa. The sawtooth waves depict the microsphere's positions at $t_1$ and $t_2$ in Fig. 1(d) to help convey this point. Additionally, we observe minima in $AF$ (blue regions) where the microsphere's position at $t_1$ is the same as at $t_2$, even if the microsphere is moving in opposite directions at those times. Note that $AF(q_r, t_1, t_2)$ is symmetric about the diagonal that runs from the top left to bottom right which denotes $\Delta t = 0$.

To determine the extent to which this pattern is maintained over the full 50 s experiment and for ROIs further from the location of the optically-trapped microsphere (which moves in a line along $y = 0$), we compare the full $AF(q_r, t_1, t_2)$ map for both an ROI close to the local strain ($y = 10.5\ \mu\text{m}$), shown in the upper right portion of Fig. 1(e), and one further away ($y = 23.5\ \mu\text{m}$), plotted in the lower left of the map. While the periodicity is obvious for $y = 10.5\ \mu\text{m}$, with distinct diamond-shaped patterns that appear to have little cycle-to-cycle variability, it is much less apparent at the distant ROI, as expected given that the degree of deformation should decay with distance from the localized perturbation site.

To better assess the dynamical anisotropy at distant ROIs, we take advantage of the well-defined periodicity of the imposed strain and perform a stroboscopic average over all cycles, which preserves the expected cycle phase dependence obvious in Fig. 1(d) and increases signal-to-noise. Specifically, we average $AF$ over all cycle periods for the set of all lag times that are less than or equal to the period, $\Delta t = t_2 - t_1 \leq T$. For speeds of $u$ = 6, 12, and 24 µm/s, this method entails averaging of 7, 15, and 30 cycles, respectively.

As shown in Fig. 2(a,b), comparison between $AF$ within the first cycle and the stroboscopic average over all cycles shows a noticeably enhanced periodic signal in the case of stroboscopic averaging, particularly in cases where the deformation of the sample is weak as in Fig. 2(b). We again observe local maxima in $AF$ at lag times $\Delta t = t_2 - t_1$ equal to half of the cycle period and bounded by the stroke extrema, as expected for largely elastic material that returns to its initial state after each loading. To better visualize changes in $AF$ throughout the cycle, we define the "cycle phase," $\phi$, based on the periodicity of $AF$. The data show rich features as a function of both lag time and phase that likely encode information about both the geometric features of the loading path within the FOV, and the structural and dynamic properties of the network. In this initial study, we focus on one limiting case where $\Delta t = T/2$, which provides a key advantage in constructing a metric for the dynamical anisotropy throughout the stroke. We define $\phi = 0$ when the intensity difference is calculated using microsphere probe positions $x(t_1) = 0$ moving in the negative *x*-direction and $x(t_2) = 0$ moving in the positive *x*-direction. Similarly, $\phi = \pi/2$ when $\Delta t = T/2$ and the microsphere probe is at $x(t_1) = -10\ \mu\text{m}$ and at $x(t_2) = +10\ \mu\text{m}$. We find similar patterns when the intensity difference is calculated using $x(t_1) = 0$ moving in the positive *x*-direction and at $x(t_2) = 0$ moving in the negative *x*-direction (corresponding to $\phi = \pi$) and $x(t_1) = +10\ \mu\text{m}$ and at $x(t_2) = -10\ \mu\text{m}$ (corresponding to $\phi = 3\pi/2$).

Importantly, with this construction, we robustly find the smallest $AF$ values when the particle passes the midplane ($\phi = 0$ or $\phi = n\pi$ where $n$ is an integer), and the largest values when $\phi \rightarrow \pi/2$ when the particle is at opposing extremal values at $t_1$ and $t_2$. This may appear counterintuitive if one expects the material deformation signal to be maximal when the particle position is closest to the midplane of motion and thus closest to the ROI's center. We attribute the small $AF$ values near the midplane to the approximate reversibility of material deformation under cyclic strain, as expected for an elastic material, and our choice of $\Delta t = T/2$. While the magnitude of material deformation may be large at this point within the cycle, the intensity difference calculated by comparing the images during the outbound and inbound portions of the cycle is small (see Eqn. 1), and thus $AF$ is near or at its minimal value. By contrast, the largest $AF$ signals are recorded when the particle pairs are maximally separated, thus accentuating the effects of the long length scale forces transmitted to the ROI. This construction offers a significant benefit: we are able to isolate the changes in material dynamics at the extremal positions of the optical trap while focusing our image analysis only on the midplane ROIs. In the context of the general starDDM framework, other choices for $\Delta t$ and other placements of the ROI are possible, allowing a broader range of material dynamics to be accessed.

In practice, to reduce the dimensionality of the data and evaluate the alignment as a function of the cycle phase $\varphi$, we limit our analysis to a central subset of lag times (half the oscillation period $\pm$ 0.21 seconds) around where maxima occur (shown via the highlighted strip on Fig. 2(a,b)). These lag times, all centered around $\Delta t = T/2$ for each speed, correspond 3.12 – 3.54 s, 1.42 – 1.84 s, and 0.58 – 1.01 s for $u = 6$, 12, and 24 μm/s, respectively. Resulting 1D average $\langle AF(\varphi)\rangle_{cycle,trials}$ functions, averaged across all trials, are shown in Fig. 2(c), with the standard deviation indicated by the shaded region. Examining $\langle AF(\varphi)\rangle_{cycle,trials}$ for different ROIs at different $y$ distances (Fig. 2(d)) clearly shows that deformations decay with distance while preserving their cyclic nature.

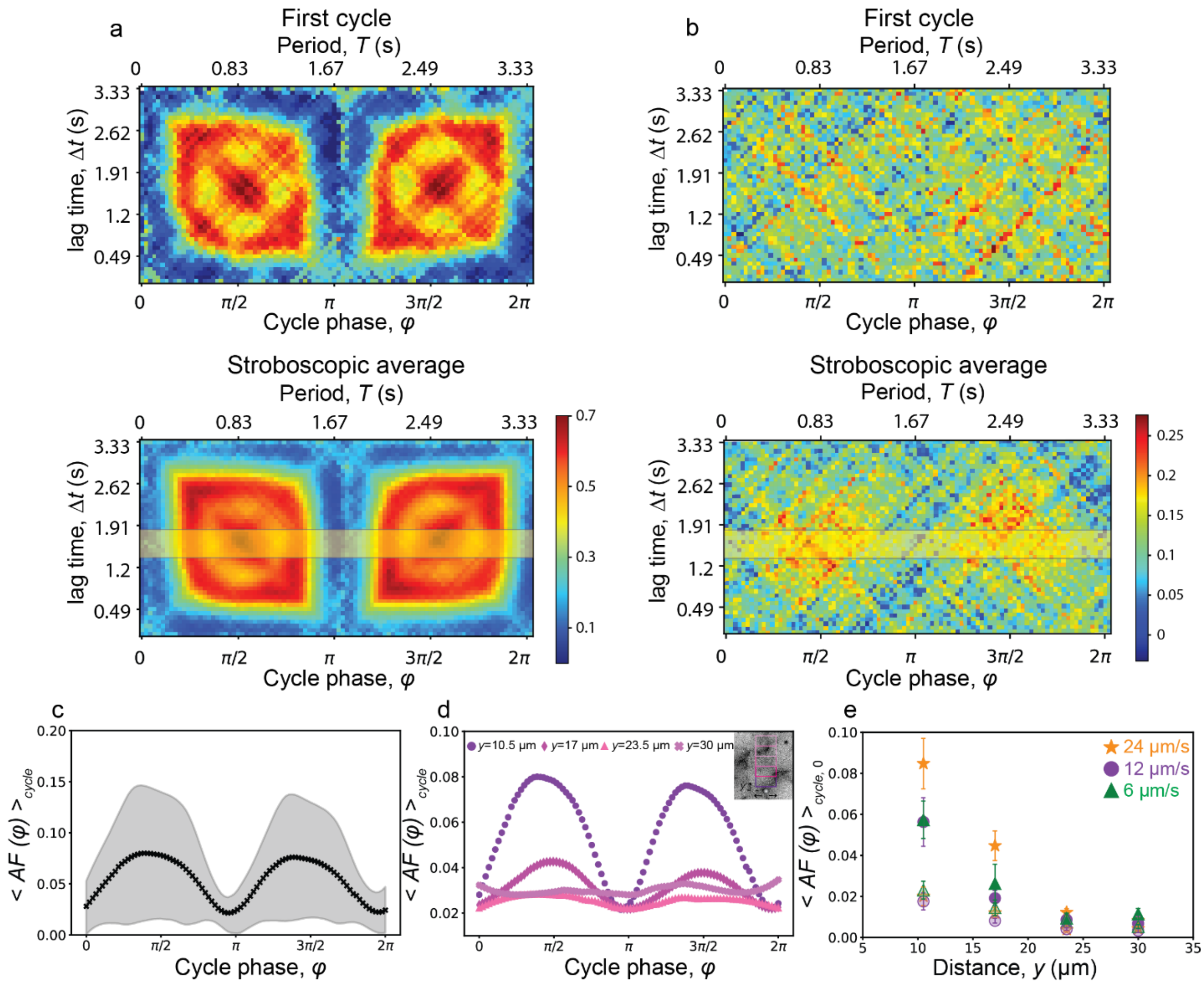

**Figure 2. Stroboscopic averaging enables characterization of probe-induced material motion from low-signal-to-noise data. (a,b)** Alignment factor values are shown as a function of time, shown here also in terms of the phase of the strain cycle, and lag time for the first cycle (top) and corresponding stroboscopic average (bottom) over 15 cycles at **(a)** $y = 10.5$ μm and **(b)** $y = 23.5$ μm. Note that the color scales are different between (a) and (b). **(c)** Averaging across a central subset of lag times $\Delta t$ = 1.418 – 1.843 s, shown as the highlighted regions of (a) and (b), we obtain $AF$ as a function of the phase $\varphi$ of the strain cycle. Here, $\langle AF(\varphi)\rangle_{cycle}$ is shown after averaging over all trials with the shaded band indicating the standard deviation. **(d)** $\langle AF(\varphi)\rangle_{cycle}$ shown for ROIs at different distances along $y$ from the local strain. ROI positions are shown in the inset. **(e)** $\langle AF\rangle_{cycle,0}$ (bold symbols) and time-averaged $\langle AF\rangle_t$ (faded symbols) both reflect the reduction in strain-induced deformation with distance, and the time-resolved $\langle AF\rangle_{cycle,0}$ values further reveal a dependence on strain speed.

Notably this strong and reproducible cyclic dependence on $\varphi$, with pronounced peaks at $\varphi \simeq \frac{\pi}{2}$ and $\varphi \simeq \frac{3\pi}{2}$, is obscured in time-averaged approaches because the single $\langle AF\rangle_t$ value for a given condition (material composition, strain speed, etc.) includes contributions from all $\varphi$ values. Such time-averaged values necessarily include those with very low signal (Fig. 1(e)), thereby reducing the signal-to-noise and the apparent magnitude of $\langle AF \rangle_t$. Conversely, in starDDM, after computing $\langle AF(\varphi) \rangle_{cycle}$, we can define what is essentially the amplitude of the time-dependent $AF$ by identifying the peak and trough values and taking their difference, which we denote as $\langle AF \rangle_{cycle,0}$. As shown in Fig. 2(e), this method amplifies the effects of the strain-dependent anisotropy of probe-induced motion and reduces the variation in each measurement. Moreover, $\langle AF \rangle_{cycle,0}$ is significantly reduced in dimensionality as compared to the full 2D map of $AF$, facilitating analysis and comparisons across conditions.

We interpret $\langle AF \rangle_{cycle,0}$ as a measure of the shear-aligned material motion, which in this construction is dominated by the difference in image intensities at the extremal positions of the optical trap during the cyclic sweep. In the case studied here ($\Delta t = T/2$, for an ROI placed above the midplane), we expect both $AF$ and $\langle AF \rangle_{cycle,0}$ to reflect the ability of the material to sustain long range material motion and dynamical anisotropy under loading. Thus, we expect more elastic materials to provide larger values of $AF$ and $\langle AF \rangle_{cycle,0}$ since such materials can sustain stress over large (~10 μm) length scales. By contrast, materials that dissipate stress, for example, through viscoelastic or plastic interactions, are expected to produce lower $AF$ and $\langle AF \rangle_{cycle,0}$ since stresses are dissipated rather than propagated through the material. This is consistent with the analytical calculations shown in Eqns. 4-7 and expanded upon in SI Section 3.

Finding the amplitude of the stroboscopically averaged $AF$ allows us to evaluate trends in material deformation across conditions (as shown in Fig. 2(c-e)) and reveals effects of probe particle speed and kinesin concentration on $\langle AF \rangle_{cycle,0}$ that were not captured with the time-averaged approach (Fig. S3, S4). However, some features in $AF$ are ignored when we consider just this amplitude. When we examine the stroboscopically averaged $AF(t, \Delta t)$ for each of the three strain speeds for $c_k = 0$ nM, a distinct square shaped pattern is observed (Fig. 3). Rather than exhibiting a prominent maximum when the image difference pair is constructed using the extremal positions of the optically trapped particle ($\phi = \frac{\pi}{2}, \frac{3\pi}{2}$), i.e., when there is the greatest probe separation distance during the cycle, as we expect for an elastic material, we observe that $AF$ fluctuates around a plateau value. This plateau appears somewhat more prominent at low speeds, which might indicate that while the networks are primarily elastic, stress dissipation plays a secondary role, consistent with prior studies of the mechanical properties of these composites.[36] This interpretation is further supported by the results of the $c_k = 0$ nM condition shown in Fig. 2(e), which demonstrate that $\langle AF \rangle_{cycle,0}$ is greatest for $u = 24$ μm/s. We attribute this to the ability of the composite to partially dissipate stress at the lower strain speeds, which has been shown to attenuate their long-range elastic response,[36] and here is shown to supress long-range motion and dynamic anisotropy of filaments within the cycle. $AF$ could also be reduced by yielding due to filament or crosslinker rupture, or the presence of nonaffine deformations that could also lead to nonlinear relationships between the motion of the optical trap and the motion and alignment of filaments within the ROI. Alternatively, it is possible that the analysis loses sensitivity in the determination of $AF$ at the largest separation distances between the particle and the ROI.

Because of the phenomenological nature of $AF$, it is not possible to precisely relate its value to a specific microscopic parameter; however, we observe trends that appear to correlate to important properties of the cytoskeletal network. In our limiting case ($\Delta t = T/2$, for an ROI placed above the midplane) we interpret network conditions producing higher values of $AF$ and $\langle AF \rangle_{cycle,0}$ as having a better ability to sustain long range material motion and alignment with the strain. With this framework in mind, we next use starDDM to investigate how the measured signal depends on sample composition (i.e., kinesin concentration, $c_k$) and strain speed $v$, which we were unable to extract with time-averaged approaches.

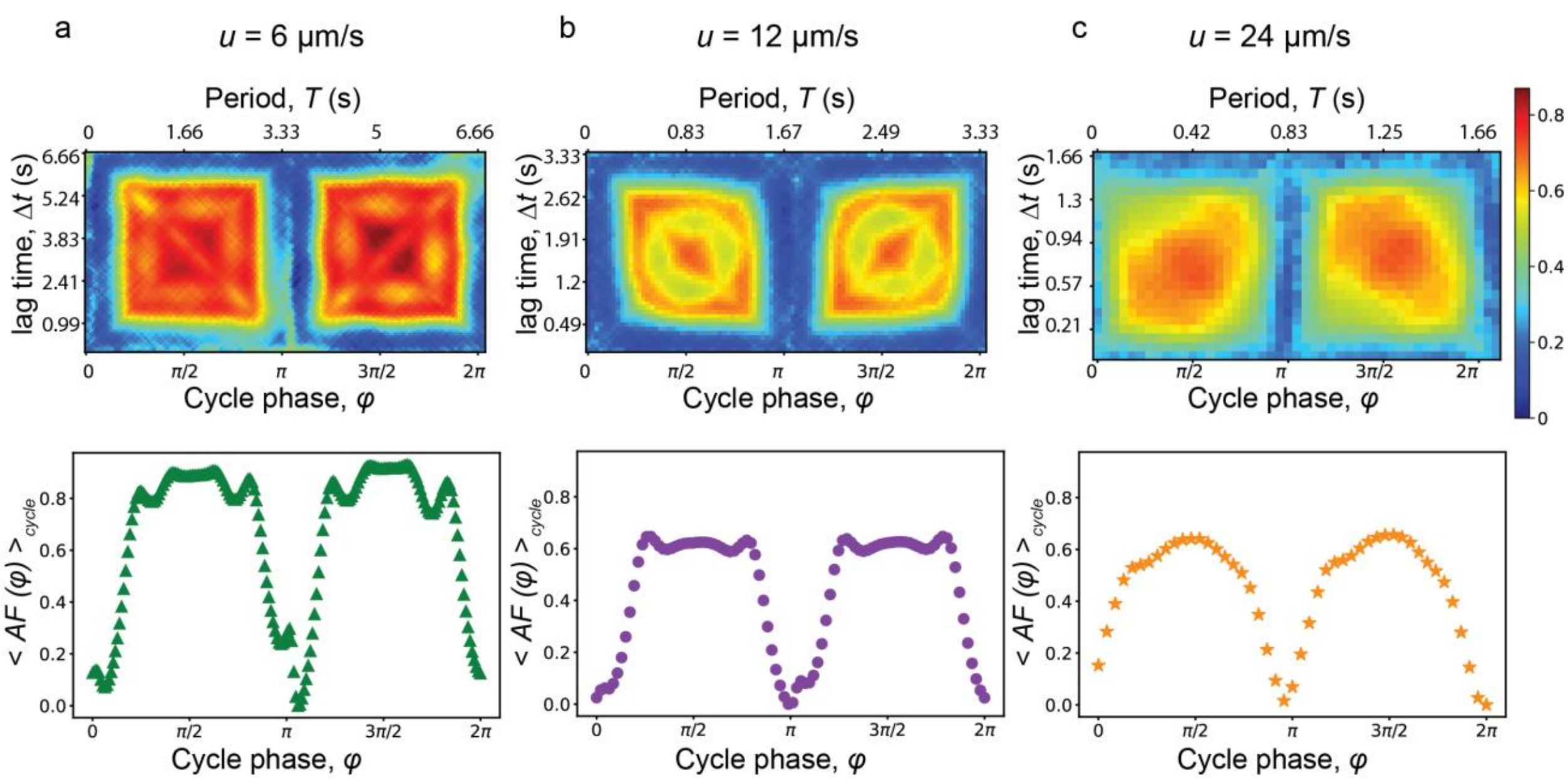

**Figure 3. Intracycle features of the alignment factor vary with strain speed.** The alignment factor for the closest ROI ($y$ = 10.5 μm) is shown following stroboscopic averaging of (**a**) 7, (**b**) 15, and (**c**) 30 cycles for strain speeds of 6, 12, and 24 $\mu m/s$, respectively. Below each heatmap of the $AF$ as a function of cycle phase and lag time, the values of $AF$ averaged over a range of lag times corresponding to half the oscillation period ± 0.21 s, as described in the text, are plotted versus the cycle phase. All data shown is for $c_k = 0$ nM.

As shown in Fig. 4(a,b,c), we observe a general increase in the alignment factor as the concentration of kinesin increases up to approximately $c_k = 160$ nM, and a plateau or decrease in $AF$ at the higher concentrations up to $c_k = 640$ nM. This trend is preserved when $\langle AF \rangle_{cycle,0}$ is averaged over all trials at the ROI closest to the applied strain, as shown in Fig. 4(d). There is considerable trial-to-trial variability in $\langle AF(\varphi) \rangle_{cycle}$ (Figs. S8, S9), and the greatest variability is observed for $c_k = 160$ nM. These results are in agreement with prior work using identical networks in which we established a non-monotonic dependence of composite elasticity on kinesin concentration due to emerging structural heterogeneity and network demixing, with a peak in stiffness at $c_k = 160$ nM.[36] This suggests that $\langle AF(\varphi) \rangle_{cycle}$ is sensitive to increases in material stiffness and connectivity, which both promote long-range transmission of stress.

To further explore the sensitivity of $AF$ to the ability for a material to transmit stresses over large length scales, we use the same ROI-based approach as above (see Fig. 1e), but now examine how $\langle AF \rangle_{cycle,0}$ varies with increasing vertical offset distance, $y$, from the axis of probe motion. For all conditions, we find a decrease in $\langle AF \rangle_{cycle,0}$ with increasing $y$, consistent with the increased separation distance between the particle path and the ROI. We identify noticeable differences in how $\langle AF \rangle_{cycle,0}$ decays with increasing $y$ for different speeds and kinesin concentrations (Fig. 5(a,b,c)) that were not resolvable using a time-averaged approach (Fig. S4). The most pronounced decrease in $\langle AF \rangle_{cycle,0}$ with distance is observed for $c_k$ = 640 nM and is present across all probe speeds (Fig. S10). This is consistent with the motor-driven demixing, loss of network connectivity, and the formation of a heterogeneous ensemble of discrete, unconnected foci previously observed, that would disrupt long-range stress propagation.[36] For intermediate motor concentrations, in which networks remain connected, we observe a less drastic decrease in $\langle AF \rangle_{cycle,0}$ indicative of more efficient stress propagation. The dominant mechanical responses are elastic in these cases, as evidenced by similar responses across multiple cycles (Fig. 1(d,e)) and $\langle AF \rangle_{cycle,0} \sim 0$ for $\phi = n\pi$, indicating largely reversible motion. However, the dependence of $\langle AF \rangle_{cycle,0}$ on $y$ depends on the particle speed, suggesting a small degree of dissipation. Specifically, the loss of $AF$ signal at the lowest speed, is likely due to the ability of the networks to relax imposed stress over this extended cycle time, which would minimize the long-range propagation of stress.

Finally, we further investigate long-range load transfer by fitting $\langle AF \rangle_{cycle,0}$ as a function of distance from the cyclically-driven optical trap, $y$, to a power law (Fig. 5(d)) for each kinesin concentration to qualitatively capture the trend. We find power law exponents <−2 in all cases, and observe steeper power laws, with exponents of −2.7 to −3.0, for the higher kinesin concentrations ($c_k \geq$160 nM) compared to the lower kinesin concentrations, with exponents within the range of −2.1 to −2.5. This is consistent with the general trends described above: motor driven demixing leads to loss of network connectivity and compromises the long-range elastic transfer of stress. Interestingly, in prior work that assessed the force response of the moving particle,[36] we observed the material to be stiffer on the initial cycle than in subsequent cycles, which we previously attributed to localized damage or plastic deformation of the pristine material as a result of the driven bead motion. The lack of such a signature in this study (demonstrated by the robust cyclic signatures observed in Fig. 1(d,e)) suggests that the responses measured at more distant ROIs may be more robust to such localized mechanical damage than those measured in close proximity to the bead.

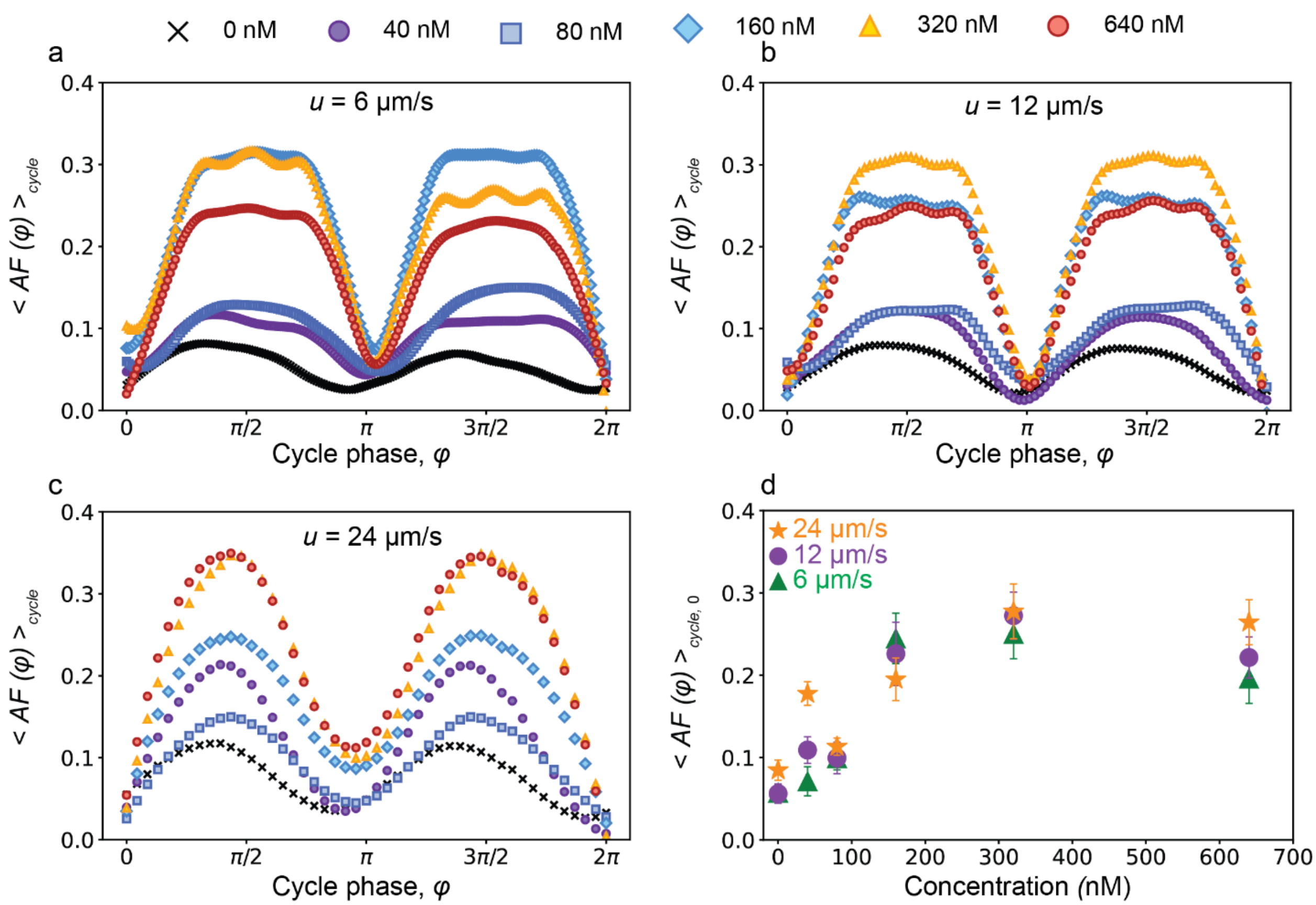

**Figure 4. Strain-induced deformation varies with kinesin concentration.** The values of $\langle AF(\varphi)\rangle_{cycle}$ are shown for all kinesin concentrations as indicated in the top legend for strain speeds of **(a)** 6 µm/s, **(b)** 12 µm/s, and **(c)** 24 µm/s, all for the closest ROI at $y = 10.5$ µm. **(d)** Variation of $\langle AF\rangle_{cycle,0}$ with kinesin concentration at $y = 10.5$ µm. For all strain speeds, we observe that the strain-induced deformations are lower for kinesin concentrations below 160 nM compared to concentrations ≥160 nM.

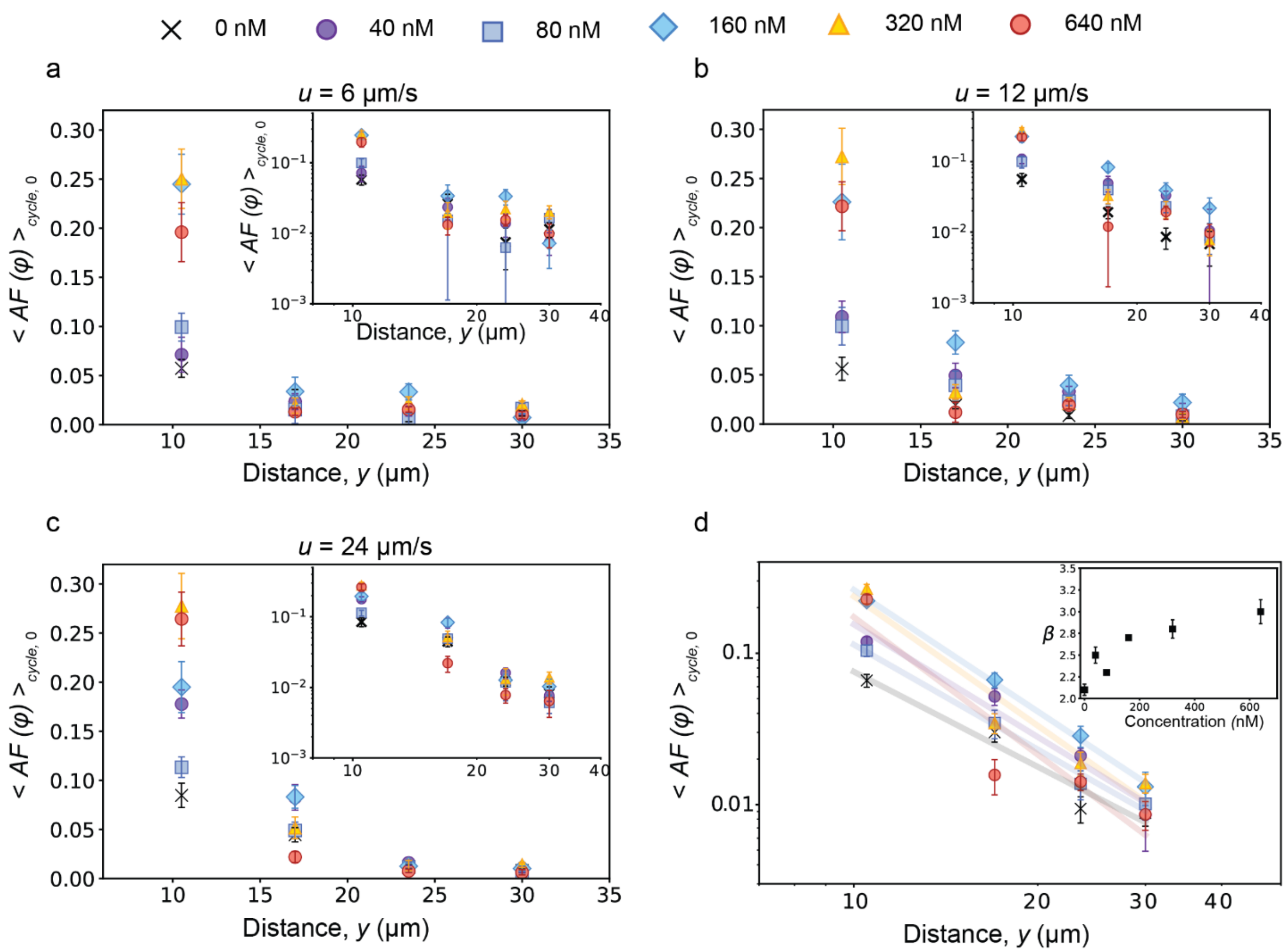

**Figure 5. The decay of $\langle AF\rangle_{cycle,0}$ with distance depends on strain speed and motor concentration.** Values of $\langle AF\rangle_{cycle,0}$ vs. distance are shown for all concentrations (as listed in the legend above the plots) for strain speeds $u$ (μm/s) of **(a)** 6, **(b)** 12, and **(c)** 24. **(d)** Decay of $\langle AF\rangle_{cycle,0}$ with distance averaged across all speeds is shown for each $c_k$. Fits of the data to a power-law function $\langle AF\rangle_{cycle,0} \sim y^{-\beta}$ yield exponents of $\beta$ = 2.1, 2.5, 2.3, 2.7, 2.8, and 3.0 for $c_k$ = 0, 40, 80, 160, 320, and 640 nM, respectively (see inset).

## Conclusion

We describe STroboscopically-Averaged time-Resolved DDM (starDDM), a method to investigate how heterogeneous soft and biological materials respond to time-periodic local perturbations. We establish a model-independent, time-resolved extension of DDM that is robust in low signal-to-noise and spatiotemporally heterogeneous conditions. Rather than modeling the time-averaged ISF and requiring measurement over timescales that capture the near complete decorrelation of density fluctuations, we avoid the use of full temporal averages, restrict analyses to small ROIs to achieve spatial resolution, and quantify the anisotropy in $D(\mathbf{q},\Delta t)$. Through stroboscopic averaging across multiple trials, and judicious choice of $\Delta t$ and ROI, we establish a relatively simple measure of strain-induced motion and anisotropy, $\langle AF \rangle_{cycle,0}$, which can be compared across conditions, speeds, and distances from the location of the applied perturbation.

We have applied this analysis to cyclically loaded kinesin-driven cytoskeleton composites, which are difficult to characterize due to the small sample volumes, small and nonuniform deformation fields, and trial-to-trial and spatial heterogeneity. With starDDM, we characterize important features and trends. For example, we show that for ROIs closest to the probe path, the strain-induced dynamical anisotropy is greater for samples with kinesin concentrations of 160 – 640 nM as compared to concentrations of 0 – 80 nM, consistent with the ability of tetrameric kinesins to crosslink and bundle microtubules, thereby enhancing local connectivity that allows for stress propagation. We find an attenuation in the magnitude of $AF$ at the highest concentrations, consistent with loss of percolation due to kinesin-driven demixing that limits the ability for the material to transmit stresses. We find trial-to-trial variability in $AF$ is highest for $c_k$ = 160 nM, consistent with the greater microscale structural heterogeneity previously identified at this kinesin concentration.

StarDDM provides a valuable addition to the available image analysis methods to characterize the response of soft materials to various external fields, that specifically focuses on discerning the strain-induced motion of the material and how the imposed deformations and stresses are transmitted over large lengthscales. Additionally, starDDM is especially useful for materials with minimal thermal fluctuations, in contrast to other DDM-based methods, as well as those with limited signal-to-noise. While we demonstrate our approach using the cyclic motion of a high-stroke optical trap, the time-resolved DDM and stroboscopic averaging approaches we describe could be used with other cyclic perturbations, caused, for example, by varying electromagnetic fields, by chemical concentrations varied with microfluidics, or by time-modulated light patterns in combination with photoresponsive materials. Future refinements to starDDM will focus on more quantitatively determining strain fields from the alignment factor and understanding how the intracycle features of $AF(\varphi)$ directly relate to conventional metrics of mechanical response, such as Poisson ratio, viscoelasticity, yielding and plasticity.

## Acknowledgments

The authors acknowledge funding from the US National Science Foundation DMREF program through following grants: NSF DMR-2119663 (to RMRA) and NSF DMR-2118497 (to MTV) with with partial support from the MRSEC Program of the National Science Foundation (NSF) under Award No. DMR 2308708 (IRG-2). The authors acknowledge the contributions of Daisy Achiriloaie in collecting experimental data and Karthik Peddireddy for acquisition software design. Google Gemini models were used in the early stages of formulation of the mathematical framework presented in SI Section 3; results presented are the output of the authors.

**Supplementary Information**

# Spatiotemporally-resolved differential dynamic microscopy reveals strain-induced deformations in heterogeneous soft materials

Aysan Razzaghi, Mehrzad Sasanpour, Jasmine Ren, Caidric Gupit, Matthew E. Helgeson, Megan T. Valentine, Rae M Robertson-Anderson*, Ryan McGorty*

**Section 1: Experimental details**

Composites of rabbit skeletal actin (Cytoskeleton, Inc. AKL99) and porcine brain tubulin (Cytoskeleton, Inc. T240) were polymerized at a 45:55 molar ratio together with with 0.1 %(v/v) of 4.5-μm-diameter carboxylated microspheres (Polysciences, Inc.). The microspheres were coated with BSA (6 $\mu M$ in borate buffer) to inhibit non-specific interactions with the network. Details of protein preparation and copolymerization are provided in Ref. 1. Kinesin motor clusters were then added to achieve final concentrations, $c_k$ = 0,40, 80, 160, 320, and 640 nM.[1]

After 30 minutes, the sample was flowed into glass-cover slip chamber with 100 μm depth that was also passivated by BSA. The chamber was sealed with UV curable glue to prevent sample leakage and evaporation. Experiments were performed using an optical trap (stiffness of 68 $pN\ \mu m^{-1}$) formed by a 1064 nm Nd:YAG fiber laser (Manlight), focused with a 60× 1.4 NA objective (Olympus), and custom-built around an Olympus IX71 epifluorescence microscope. Imaging of the labeled filaments and probes was achieved using a broadband LED source (XCITE) with 488nm/525nm excitation/emission filters and a Hamamatsu ORCA-Flash 4.0LT CMOS camera at a frame rate of 20 fps. The trapped microsphere was moved back and forth through the sample using a piezoelectric actuator mirror (PI USA) while keeping the sample chamber fixed. The total perturbation time was fixed at 50 seconds with the stage movement of 20 μm per half a cycle. Measurements were performed at 3 different speeds for each kinesin concentration: $u$ = 6, 12, 24 μm/s that imposes shear rate of $\dot{\gamma}$ = 5.7, 11.3, and 22.6 $s^{-1}$, respectively. For each $(u, c_k)$ combination, measurements were repeated using multiple beads located in different regions of the sample chamber, which were separated by ≥100 μm. This suite of measurements for each condition was repeated for three different samples for a total of $\geq 20$ beads per condition.

**Section 2: Evaluating the source of DDM matrix asymmetry**

First, we evaluated if complete decorrelation of density fluctuations within a ROI in a particular direction contributed to the asymmetry in the DDM matrix as measured through the alignment factor. For example, if the motion of the network in the direction of the strain ($x$) was strong enough for density fluctuations along that direction at a particular $q$ and lag time to completely decay, then the intermediate scattering function along that direction (ISF) would decay to zero. This would have the potential of making comparisons of $AF$ across conditions or distances from the strain less meaningful. We tested if this were the case by performing sector averaging, rather than full azimuthal averaging, of the DDM matrix. To perform sector averaging in Fourier space, angular sectors centered along the $q_x$ and $q_y$ directions were defined, each spanning $\pm 7.5°$. Within these sectors, the DDM matrix was azimuthally averaged over the corresponding range of $q_r$. The amplitude, $A(q_r)$, and background, $B$, were then determined to find the intermediate scattering function (ISF) from $D(q_r)$. The background was assumed to be independent of $q$ and estimated as the mean value over all lag times of the DDM matrix at the highest measured $q$-value. In other words, $B = \langle D(q_{max}, \Delta t) \rangle_{\Delta t}$. To find the amplitude term, $A(q_r)$, we assumed that $A(q_r) + B = 2 \langle \left| \tilde{I}(q_r) \right|^2 \rangle$. The ISF was then calculated as:

$$\mathrm{ISF} = 1 - \frac{D(q_r, \Delta t) - B}{A(q_r)}. \tag{S1}$$

Figure S1 shows the ISF as a function of lag time along the shear direction (*x*) and perpendicular to it (*y*). In neither direction does the ISF decay to zero at the lag times used to calculate the alignment factor, suggesting that the alignment factor values we calculate are not being clipped by complete decorrelations.

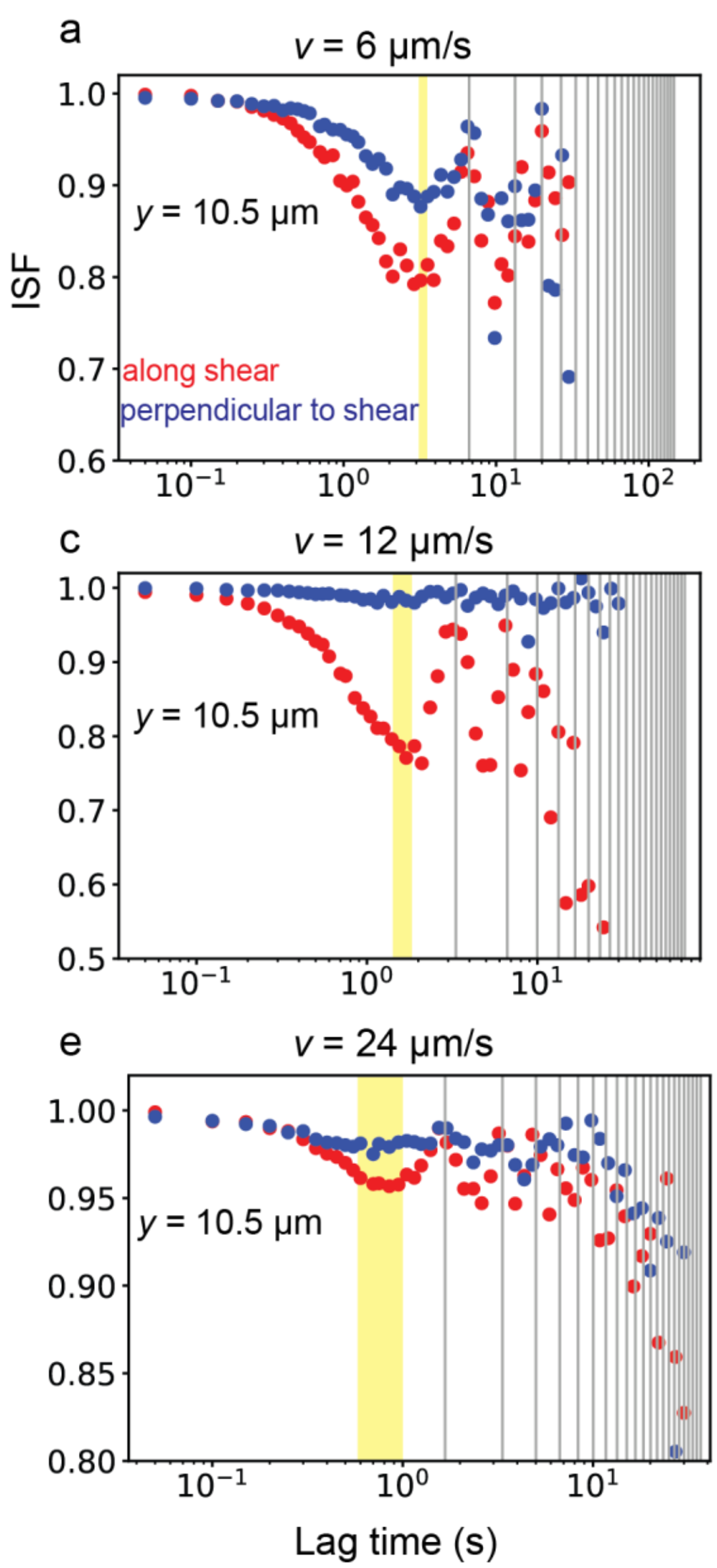

**Figure S1**. The intermediate scattering function (ISF) is plotted as a function of lag time at $q$ = 1.5 μm$^{-1}$. In red and blue we plot the ISF along the *x*- and *y*-directions, respectively, determined by considering a sector of $\pm 7.5°$ in $q$-space, as described above. The range of lag times highlighted in yellow correspond to what we consider when calculating the alignment factor. Periodicity of the optically trapped microsphere ($\tau = X/u$, where $X$ is 40 $\mu$m) is shown with vertical gray lines. At each speed, the ISFs are shown for the closest ROI ($y$ = 10.5 μm) and the third ROI ($y$ = 23.5 μm). The microspheres were moved at (a,b) $u$ = 6 μm/s, (c,d) $u$ = 12 μm/s, and (e,f) $u$ = 24 μm/s.

Second, we evaluated whether the asymmetry of the DDM matrix could be the result of anisotropy in structure rather than dynamics. Recall that the DDM matrix can be modeled as $D(\mathbf{q}, \Delta t) = A(\mathbf{q})\{1 - f(\mathbf{q}, \Delta t)\} + B(\mathbf{q})$ where $A(\mathbf{q})$ depends on the structure of objects and properties of the optical systems, $f(\mathbf{q}, \Delta t)$ is the intermediate scattering function (ISF) that describes the dynamics, and $B(\mathbf{q})$ is the background term. Anisotropy in the structure, for example, filaments that preferentially align along *x* or *y*, would result in anisotropy in $A(\mathbf{q})$ and, therefore, in $D(\mathbf{q}, \Delta t)$, even if the dynamics were isotropic.

Since the power spectrum of the images (rather than of the difference between two images separated by some lag time) is proportional to $A(\mathbf{q})$, but not to $f(\mathbf{q}, \Delta t)$, we calculated this quantity, $\langle |\tilde{I}(q_r, t, \theta)^2| \rangle$. We then calculated the anisotropy of this term analogously to how we determined $AF$ from $D(\mathbf{q}, t, \Delta t)$. To differentiate this term from $AF$, we refer to it as $IAF$.

$$IAF = \int_0^{2\pi} |\tilde{I}(q_r, t, \theta)^2| \cos(2\theta) \, d\theta \Big/ \int_0^{2\pi} |\tilde{I}(q_r, t, \theta)^2| \, d\theta \quad \text{(S2)}$$

Similarly to how we found $\langle AF(\varphi) \rangle_{cycle}$ through stroboscopic averaging, we calculate $\langle IAF(\varphi) \rangle_{cycle}$. Figure S2 shows the stroboscopically averaged image alignment factor, $\langle IAF(\varphi) \rangle_{cycle}$ , for composites of different kinesin concentrations. For comparison, $\langle AF(\varphi) \rangle_{cycle}$ is shown on the right. While some cycle-phase-dependent trends are observed, we do not observe the same large amplitude variations of $\langle IAF(\varphi) \rangle_{cycle}$ with the strain cycle as we do with $\langle AF(\varphi) \rangle_{cycle}$, and do not consistently find maximal values at $\pi/2$ and $3\pi/2$. This suggests that the $\varphi$-dependence of $\langle AF(\varphi) \rangle_{cycle}$ is not solely due to variations in the structure (e.g., filaments aligning their orientation with the shear direction) but that there is a significant, and perhaps dominant, effect from the dynamic motion of the filaments.

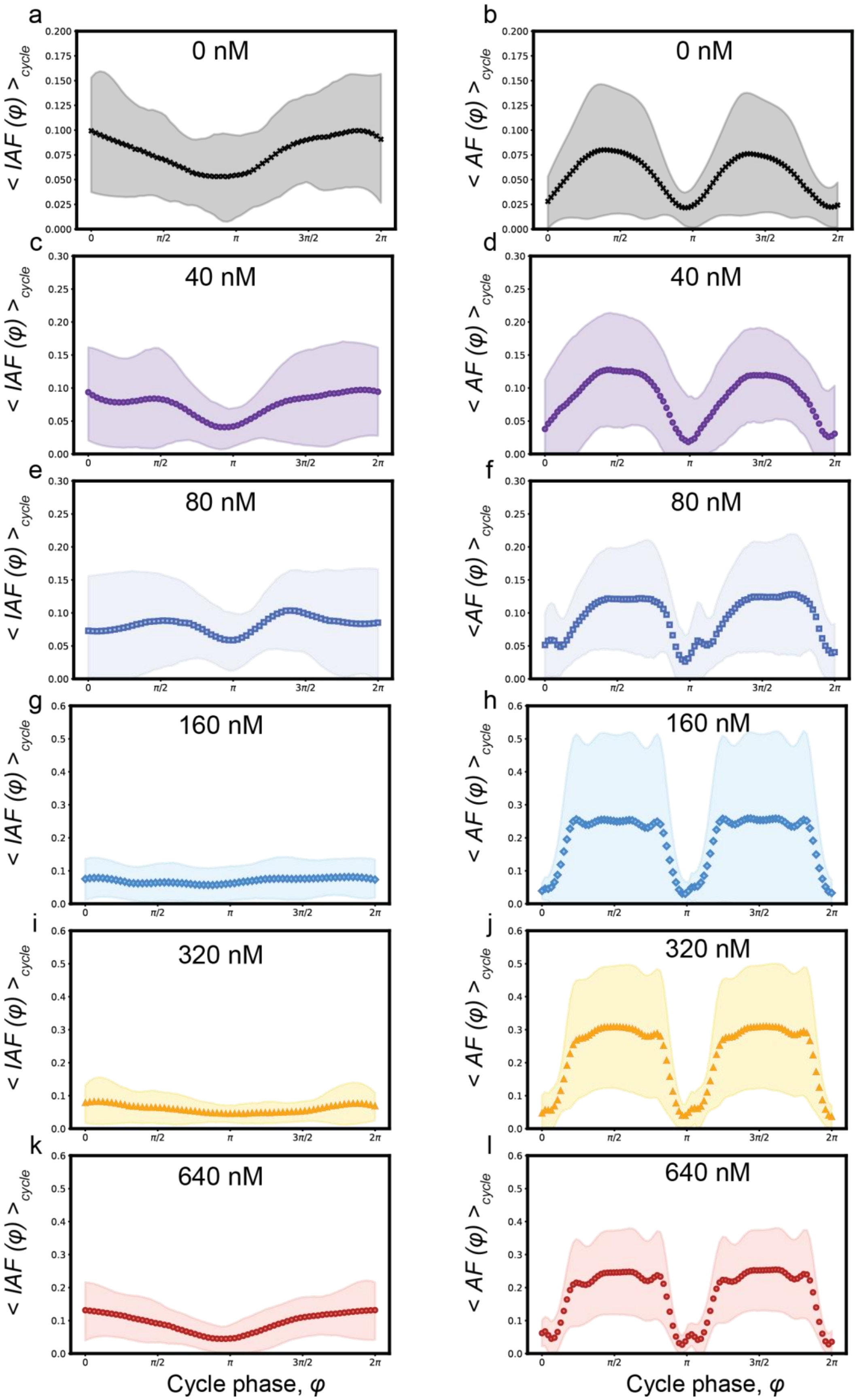
a
0 nM
< IAF (φ) >cycle
b
0 nM
< AF (φ) >cycle
c
40 nM
d
40 nM
e
80 nM
f
80 nM
<AF (φ) >cycle
g
160 nM
h
160 nM
i
320 nM
j
320 nM
k
640 nM
l
640 nM
0
π/2
π
3π/2
2π
Cycle phase, φ

**Figure S2**. Comparison between $\langle IAF(\varphi)\rangle_{cycle}$ obtained from image power spectrum (a, c, e, g, I, and k) and $\langle AF(\varphi)\rangle_{cycle}$ (b, d, f, h, j, l) at (a, b) 0 nM kinesin, (c, d) 40 nM kinesin, (e, f) 80 nM kinesin, (g, h) 160 nM kinesin, (i, j) 320 nM kinesin, and (k, l) 640 nM kinesin concentration. The shaded are uncertainty determined based on the standard deviations in $\langle IAF(\varphi)\rangle_{cycle}$ and $\langle AF(\varphi)\rangle_{cycle}$ across 20+ trials.

## Section 3: Theory of starDDM for purely advective motion in a homogeneous medium.

### i. Two-time scattering function and anisotropy factor for affine motion in a continuum

The two-time DDM anisotropy factor is defined as

$$AF(q_r, t_1, t_2) = \frac{\int_0^{2\pi} D\,(\mathbf{q}, t_1, t_2)\cos(2\theta)\,d\theta}{\int_0^{2\pi} D\,(\mathbf{q}, t_1, t_2)\,d\theta} \tag{S3}$$

Assuming, as described in the main text, that the amplitude term $A(\mathbf{q})$ within $D(\mathbf{q}, t_1, t_2)$ is isotropic with no contribution to the measured $AF$ signal from static anisotropy of the material microstructure over the probed $q$-range, we can write $A(\mathbf{q}) = A(q_r)$. We further assume that $A(q_r) \gg B(\mathbf{q})$ which allows us to relate $AF$ to the intermediate scattering function as

$$AF(q_r, t_1, t_2) = -\frac{\frac{1}{2\pi}\int_0^{2\pi} f\,(\mathbf{q}, t_1, t_2)\cos(2\theta)\,d\theta}{1 - \frac{1}{2\pi}\int_0^{2\pi} f\,(\mathbf{q}, t_1, t_2)\,d\theta} \tag{S4}$$

where $f(\mathbf{q}, t_1, t_2)$ is the two-time intermediate scattering function. In general, $f(\mathbf{q}, t_1, t_2)$ will have contributions from both affine (advective) and non-affine (fluctuating) motions. We assume these contributions are linearly separable,

$$f(\mathbf{q}, t_1, t_2) = C f_a(\mathbf{q}, t_1, t_2) + (1 - C)\, f_f(\mathbf{q}, t_1, t_2) \tag{S5}$$

where $C$ is a linear coefficient expressing the relative contributions of the two types of motion. Further assuming that non-affine motion is negligible (i.e., $C = 1$), Eq. (S4) reduces to

$$AF(q_r, t_1, t_2) = -\frac{\frac{1}{2\pi}\int_0^{2\pi} f_a\,(\mathbf{q}, t_1, t_2)\cos(2\theta)\,d\theta}{1 - \frac{1}{2\pi}\int_0^{2\pi} f_a\,(\mathbf{q}, t_1, t_2)\,d\theta} \tag{S6}$$

In the continuum limit, averaging over an ROI of macroscopic volume $V$ within an incompressible material, the affine intermediate scattering function will depend on the spatiotemporally-varying velocity field $\mathbf{v}(\mathbf{r}, t)$ and is given by

$$f_a(\mathbf{q}, t_1, t_2) = \frac{1}{V}\int \exp\left[i\mathbf{q} \cdot \int_{t_1}^{t_2} \mathbf{v}\,(\mathbf{r}(t'|\mathbf{r}'), t')\,dt'\right] d^3\mathbf{r}'. \tag{S7}$$

Here, $\mathbf{r}(t'|\mathbf{r}')$ is the Lagrangian path function describing the time-dependent location of a material element initially at $\mathbf{r}'$ subjected to the velocity field $\mathbf{v}(\mathbf{r}, t)$.

The structure of Eq. (S7) shows that $f_a(\mathbf{q}, t_1, t_2)$ has a very complex dependence on time for a general material, the details of which are beyond the current scope. However, the integral in Eq. (S7) can be evaluated analytically in the small-deformation limit for several prototypical cases, providing initial expectations for how $AF(q_r, t_1, t_2)$ should behave for the present active microrheology experiments, in which a probe particle is displaced with respect to a fixed ROI in a homogeneous viscoelastic material by an optical trap.

### ii. starDDM for a probe moving in a viscoelastic material

We consider the behavior of Eq. (S7) in the context of active microrheology, in which a probe particle is moved through a material with a constant velocity. The probe is centered at position $\mathbf{r}^*(t)$, where $\mathbf{r}^*(t)$ moves with constant velocity $u$ beginning arbitrarily far from an ROI centered at the origin, producing a local (Lagrangian) strain displacement field $\boldsymbol{\delta}(\boldsymbol{\rho}, t)$, where $\boldsymbol{\rho} = \mathbf{r} - \mathbf{r}^*(t)$ is the local coordinate relative to the probe particle position.

Considering the case where the material's viscoelastic properties are homogeneous and time-stationary (i.e., its local viscoelastic dynamics do not depend on the position $\mathbf{r}$ or observation time *t*) produces an interesting and useful result for the intermediate scattering function. Specifically, because the Eulerian velocity field $\mathbf{v}(\mathbf{r}, t)$ is simply the time derivative of the Lagrangian displacement field, $\mathbf{v}(\boldsymbol{\rho}, t) = \partial\boldsymbol{\delta}(\boldsymbol{\rho}, t)/\,\partial t$, the two-time integral in the intermediate scattering function evaluates directly to the difference in the Lagrangian displacements at the two observation times:

$$\int_{t_1}^{t_2} \mathbf{v}\left(\mathbf{r}(t'|\mathbf{b}), t'\right) dt' = \boldsymbol{\delta}(\boldsymbol{\rho}(t_2), t_2) - \boldsymbol{\delta}(\boldsymbol{\rho}(t_1), t_1) \tag{S8}$$

so that

$$f_a(\mathbf{q}, t_1, t_2) = \exp\{i\mathbf{q} \cdot [\boldsymbol{\delta}(\boldsymbol{\rho}(t_2), t_2) - \boldsymbol{\delta}(\boldsymbol{\rho}(t_1), t_1)]\}. \tag{S9}$$

We note that this result does not depend on the particular local structure of $\boldsymbol{\delta}(\boldsymbol{\rho}, t)$, nor the chosen ROI. Thus, the intermediate scattering function is directly proportional to the Fourier mode decomposition of the Lagrangian-frame difference in local strain between the observation times $t_2$ and $t_1$.

Decomposing $\mathbf{q}$ and $\boldsymbol{\delta}$ into components parallel and perpendicular to $\boldsymbol{u}$, we obtain

$$f_a(\mathbf{q}, t_1, t_2) = \exp(i)\{\Delta\delta_{\parallel}(\boldsymbol{\rho}, t_1, t_2)\mathbf{q}_{\parallel} - \Delta\delta_{\perp}(\boldsymbol{\rho}, t_1, t_2)\mathbf{q}_{\perp}\}. \tag{S10a}$$

where $\Delta\delta_i(\boldsymbol{\rho}, t_1, t_2) = \delta_i(\boldsymbol{\rho}(t_2), t_2) - \delta_i(\boldsymbol{\rho}(t_1), t_1)$. Equivalently, in radial polar coordinates

$$f_a(\mathbf{q}, t_1, t_2) = \exp\{iq_r[\Delta\delta_{\parallel}(\boldsymbol{\rho}, t_1, t_2)\cos\theta - \Delta\delta_{\perp}(\boldsymbol{\rho}, t_1, t_2)\sin\theta]\}. \tag{S10b}$$

Substituting this expression into Eq. (S6) and simplifying yields

$$AF(q_r, t_1, t_2) = \left(\frac{\Delta\delta_{\parallel}(\boldsymbol{\rho}, t_1, t_2)^2 - \Delta\delta_{\perp}(\boldsymbol{\rho}, t_1, t_2)^2}{\Delta\delta_{\parallel}(\boldsymbol{\rho}, t_1, t_2)^2 + \Delta\delta_{\perp}(\boldsymbol{\rho}, t_1, t_2)^2}\right) \frac{J_2\left(q_r\sqrt{\Delta\delta_{\parallel}(\boldsymbol{\rho}, t_1, t_2)^2 + \Delta\delta_{\perp}(\boldsymbol{\rho}, t_1, t_2)^2}\right)}{1 + J_0\left(q_r\sqrt{\Delta\delta_{\parallel}(\boldsymbol{\rho}, t_1, t_2)^2 + \Delta\delta_{\perp}(\boldsymbol{\rho}, t_1, t_2)^2}\right)} \tag{S11}$$

where $J_0(\dots)$ and $J_2(\dots)$ are the standard Bessel functions of the first kind of order zero and two, respectively. Thus, we observe that, for the case of a random isotropically structured material undergoing affine motion in response to a probe driven with constant velocity $\boldsymbol{u}$, the anisotropy factor is expected to depend on the local strain anisotropy around the probe. We note that in cases where $\Delta\delta_{\perp} = 0$ and the total displacement is small, the right-hand side of Eq. (S11) reduces to $(q_r^2/16)\Delta\delta_{\parallel}(\boldsymbol{\rho}, t_1, t_2)^2$.

### iii. Constant-Velocity Probe in a Hookean Solid

As a toy case that provides notation and illustrates the mathematical approach for evaluating $f_a(\mathbf{q}, t_1, t_2)$ and $AF(q_r, t_1, t_2)$ for the case of a constant probe velocity in the small-strain limit, we consider an infinitesimal point probe with constant velocity $\boldsymbol{u}$ moving in a homogeneous linear elastic (Hookean) solid, for which the local displacement field is given by the Kelvin solution,

$$\boldsymbol{\delta}(\boldsymbol{\rho}, t) = \frac{1}{16\pi\mu(1-\nu)}\left(\frac{(3-4\nu)\mathbf{F}_0(t)}{|\boldsymbol{\rho}|} + \frac{(\mathbf{F}_0(t)\cdot\boldsymbol{\rho})\boldsymbol{\rho}}{|\boldsymbol{\rho}|^3}\right) \tag{S12}$$

where $\mu$ is the shear modulus, $\nu$ is the Poisson ratio, $\boldsymbol{\rho} = \mathbf{r} - \mathbf{r}^*(t)$ and $\mathbf{r}^*(t) = \boldsymbol{u}(t - t_0)$ as defined previously, and $\mathbf{F}_0(t)$ is the time-dependent force required to maintain the probe at velocity $\boldsymbol{u}$. For the case of active microrheology using an optically-driven probe, $\mathbf{F}_0(t)$ is typically measured experimentally through analysis of the relative positions of the probe and the driving optical trap. Note that in our experiments, we do not know $\mathbf{F}_0(t)$.

Under the assumption that the probe particle begins and ends far from the ROI, the total displacement $\Delta\mathbf{r}$ of material in the ROI can be evaluated on the interval $[t_1, t_2] = [-\infty, +\infty]$:

$$\Delta r = \int_{-\infty}^{+\infty} \mathbf{v}\,(\mathbf{r}(t), t)\,dt \tag{S13}$$

In the case of Eq. (S12), the problem cannot be evaluated for the case of an ROI placed precisely at the origin due to the discontinuity at $\boldsymbol{\rho} = 0$, and so we instead evaluate a case where the ROI is shifted by a vector $\mathbf{b}$ from the point of closest approach of the probe, so that $\mathbf{r}^*(t) = \boldsymbol{u}(t - t_0) + \mathbf{b}$.

With this definition, the time-dependent strain field can be rewritten solely in terms of $\mathbf{F}_0$ and $\boldsymbol{u}$ as

$$\boldsymbol{\delta}(\boldsymbol{\rho}, t) = \frac{1}{16\pi\mu(1-\nu)}\left(\frac{(3-4\nu)\mathbf{F}_0(t)}{R(t)} + \frac{\big(\mathbf{F}_0(t)\cdot(\boldsymbol{u}t+\mathbf{b})\big)(\boldsymbol{u}t+\mathbf{b})}{R(t)^3}\right) \tag{S14}$$

where $R(t) = \sqrt{u^2t^2 + b^2}$ is the time-dependent distance from the probe to the ROI. For the case considered in this work, where $\mathbf{F}_0$ and $\boldsymbol{u}$ are colinear and $\mathbf{b}$ is orthogonal to $\boldsymbol{u}$, the expression simplifies to

$$\boldsymbol{\delta}(\boldsymbol{\rho}, t) = \frac{F_0(t)}{16\pi\mu(1-\nu)}\left\{\left(\frac{(3-4\nu)}{R(t)} + \frac{u^2t^2}{R(t)^3}\right)\hat{\boldsymbol{u}} + \frac{ut}{R(t)^3}\mathbf{b}\right\} \tag{S15}$$

Substituting Eq. (S15) for the case of an ROI perpendicular to the probe path into Eq. (S10a) gives the exact relation for a probe moving in a uniform elastic medium with constant velocity,

$$f_a(\mathbf{q}, t_1, t_2) = \exp\left\{\frac{F_0(t)u}{16\pi\mu(1-\nu)}\, i\left[\frac{(3-4\nu)}{u}\left(\frac{1}{R(t_2)} - \frac{1}{R(t_1)}\right) + u_0\left(\frac{t_2^2}{R(t_2)^3} - \frac{t_1^2}{R(t_1)^3}\right)\right]\mathbf{q}_{\parallel} + \left[\frac{bt_2}{R(t_2)^3} - \frac{bt_1}{R(t_1)^3}\right]\mathbf{q}_{\perp}\right\} \tag{S16}$$

This relation corresponds only to the loading portion of the half-cycle. However, assuming the probe begins and ends a half-cycle very far from the ROI such that material points in the ROI return to their initial positions, $f_a(\mathbf{q}, t_1, t_2)$ for the full cycle will be two-time symmetric about the half-cycle, i.e.,

$$f_a(\mathbf{q}, -t_1, -t_2) = f_a(\mathbf{q}, t_1, t_2) \tag{S17}$$

when $t = 0$ corresponds to the point in the cycle where the probe motion reverses.

Ignoring the time and spatial dependence, we see that for the case of a homogeneous linear elastic medium, in addition to the strain anisotropy (represented here by the relative magnitudes of the $\mathbf{q}_{\parallel}$ and $\mathbf{q}_{\perp}$ terms), the magnitude of $AF(q_r, t_1, t_2)$ is set by the prefactor on the exponential,

$$\ln AF \propto \frac{F_0(t)\, u}{16\pi\mu(1-\nu)}. \tag{S18}$$

In terms of the time-dependence, because $\boldsymbol{\delta}(\boldsymbol{\rho}, t) \to 0$ as $t \to \pm\infty$, we obtain $f_a(\mathbf{q}, t_1, t_2) \to 1$ for times when the probe is very far from the ROI, and so $AF \to 0$ at the beginning and end of a cycle.

## Section 4: Additional figures comparing the time-averaged and time-resolved alignment factors, $\langle AF \rangle_t$ and $\langle AF(\varphi) \rangle_{cycle,0}$.

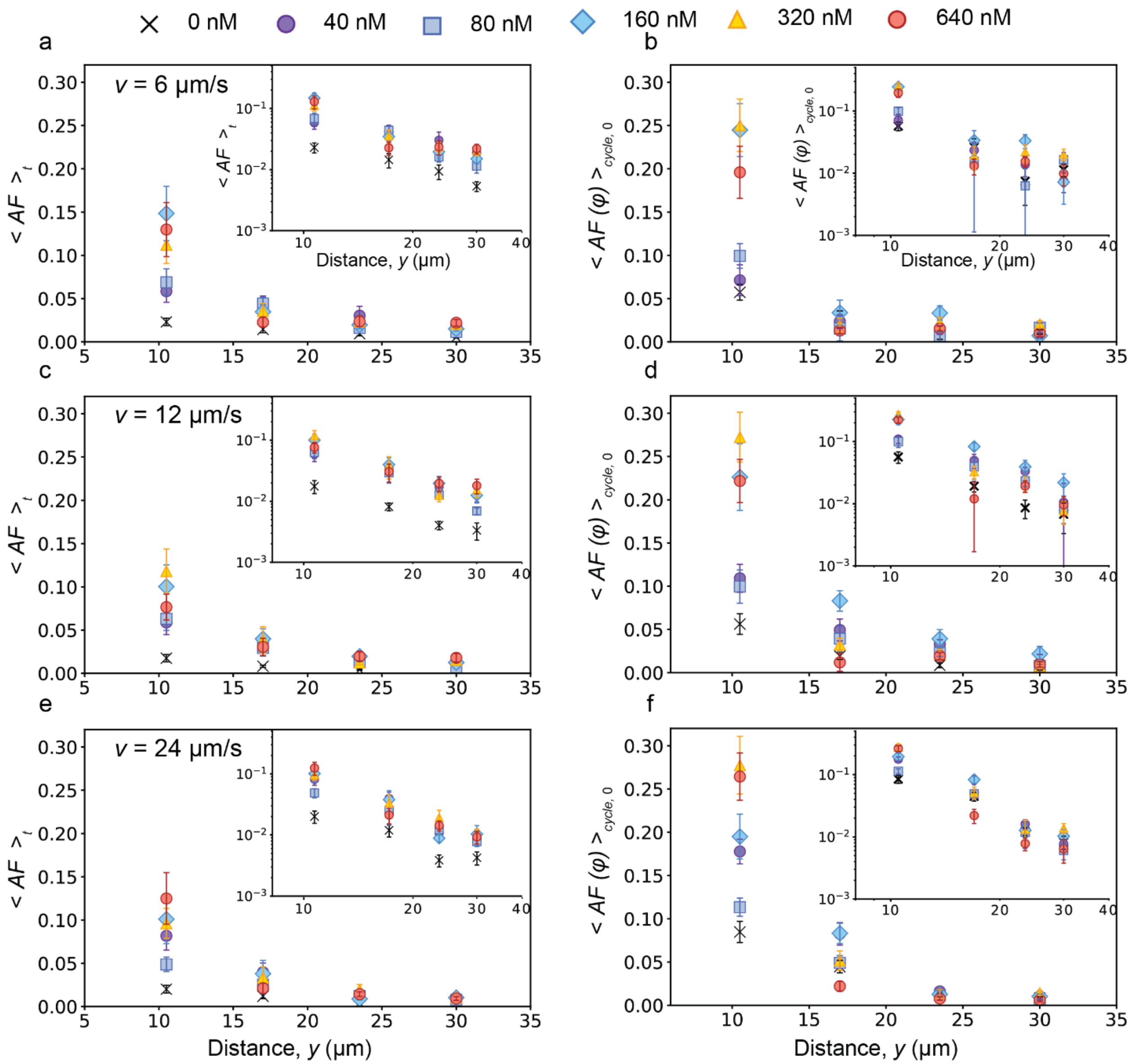

**Figure S3**. Alignment factors are plotted versus distance for each strain velocity as determined after (a,c,e) time-averaging to obtain $\langle AF \rangle_t$ and (b,d,f) using a time-resolved approach to obtain $\langle AF(\varphi) \rangle_{cycle,0}$. For each plot, all kinesin concentrations are shown as indicated in the top legend. While both the time-averaged and time-resolved approaches show a decrease of the alignment factor with distance, the time-resolved approach results in a greater than of values.

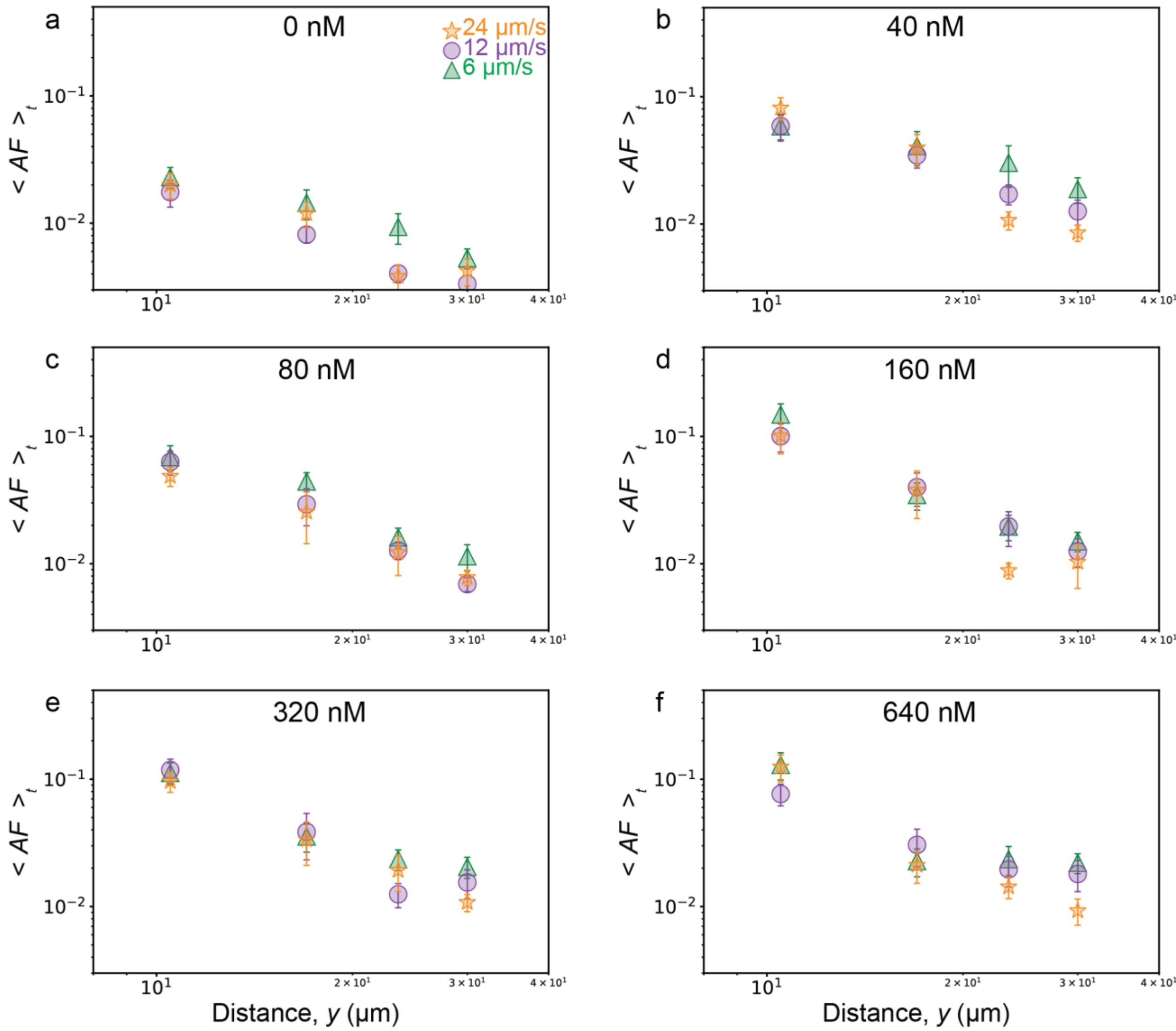

**Figure S4.** Effects of speed on the alignment factor, $\langle AF \rangle_t$, obtained from time-averaged approach. Green triangle symbols represent 6 μm/s, purple circles represent 12 μm/s, and orange stars represent 24 μm/s. While the expected decrease in alignment factors with distance is evident, differences in how the alignment factor decreases with distance for different speeds cannot be detected. Data here can be compared with the time-resolved approach of finding $\langle AF(\varphi) \rangle_{cycle}$ in Fig. S10.

## Section 5: Additional time-resolved alignment factor data

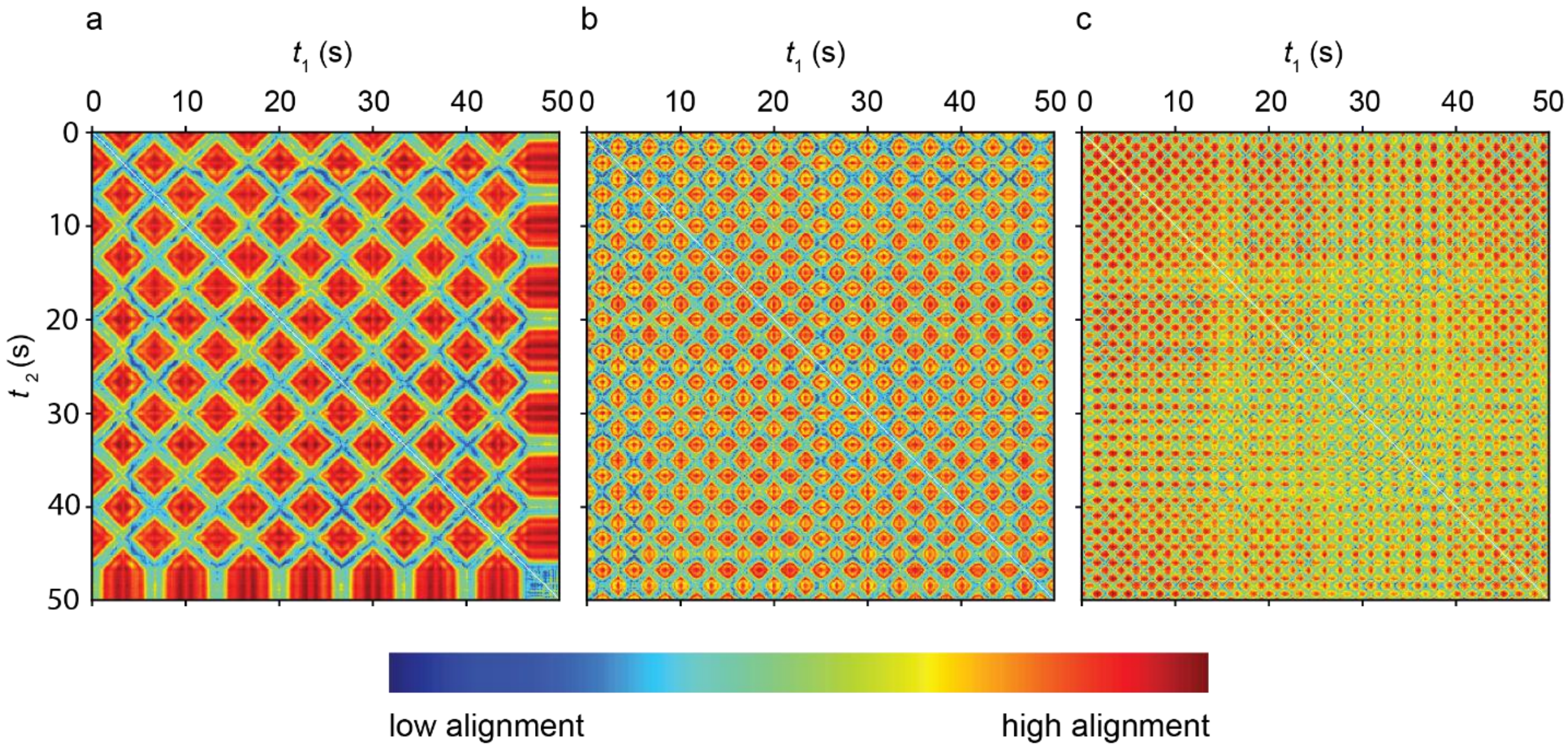

**Figure S5.** The time-resolved alignment factors $AF(q_r = 1.5\ \mu\text{m}^{-1}, t_1, t_2)$ are shown for representative movies at each of the three microsphere strain speeds: (a) 6 μm/s, (b) 12 μm/s and (c) 24 μm/s. For each, the data corresponds to the closest ROI, $y$ = 10.5 μm. The number of strain cycles contained within one 50-s movie increases with increasing strain speed.

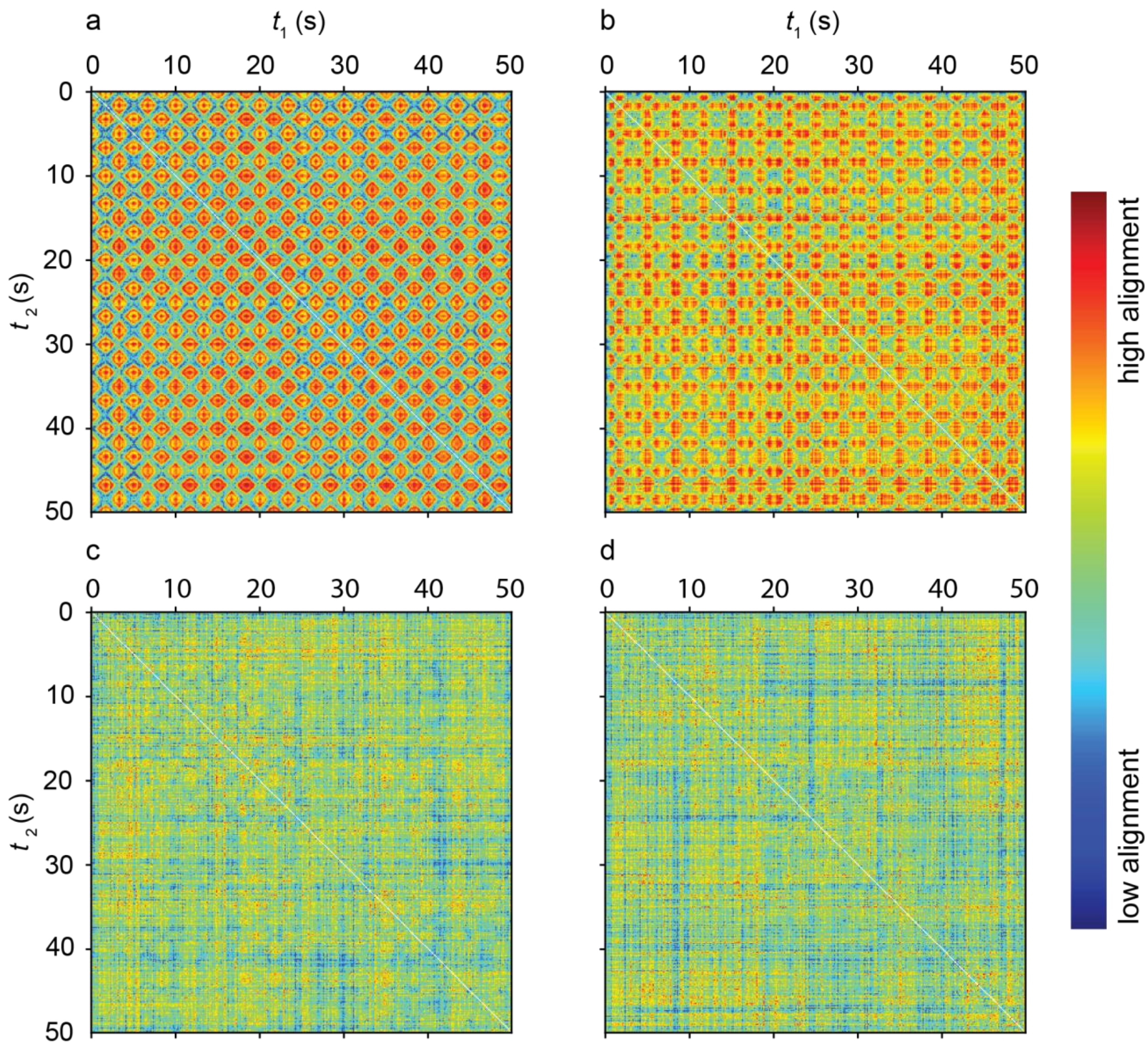

**Figure S6.** Time-resolved alignment factors, $AF(q_r = 1.5\ \mu m^{-1}, t_1, t_2)$, are shown corresponding to the entire 50 seconds of a representative movie for ROIs at (a) $y$ = 10.5 μm, (b) $y$= 17 μm, (c) $y$= 23.5 μm, and (d) $y$= 30 μm, at strain speed of 12 μm/s and kinesin concentration of $c_k = 160$ nM. At ROIs further away from the location of the moving optical trap, the periodic pattern becomes less apparent as the deformation magnitude decays with distance.

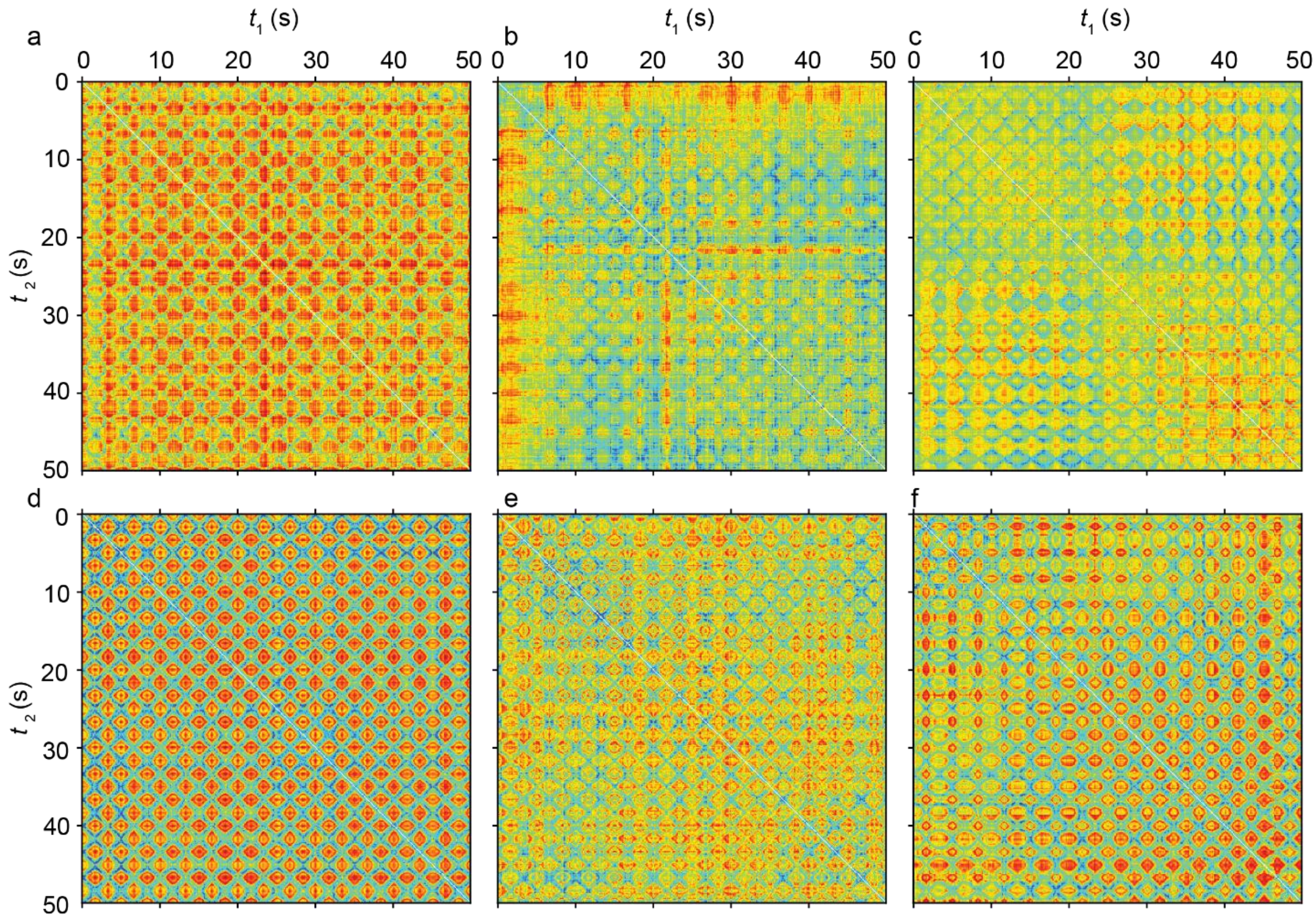

**Figure S7**. Time-resolved alignment factors, $AF(q_r = 1.5\ \mu m^{-1}, t_1, t_2)$, are shown corresponding to the entire 50 seconds of representative movies for each kinesin concentration. For each condition, data corresponds to the ROI closest to the microsphere, $y$ = 10.5 μm, and a strain speed of 12 μm/s. Kinesin concentrations are (a) 0 nM, (b) 40 nM, (c) 80 nM, (d) 160 nM, (e) 320 nM, and (f) 640 nM.

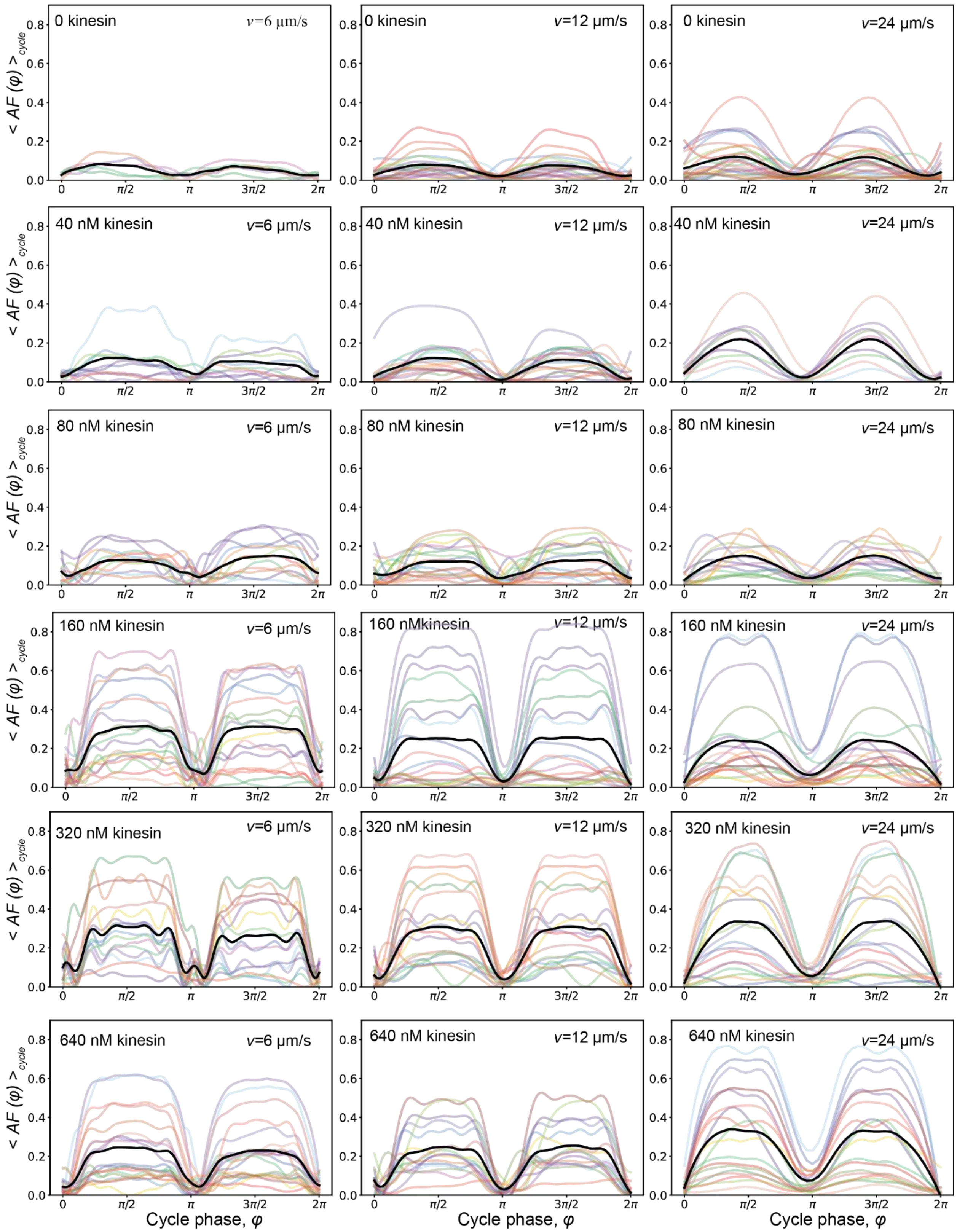

0 kinesin
v=6 µm/s
0 kinesin
v=12 µm/s
0 kinesin
v=24 µm/s
40 nM kinesin
v=6 µm/s
40 nM kinesin
v=12 µm/s
40 nM kinesin
v=24 µm/s
80 nM kinesin
v=6 µm/s
80 nM kinesin
v=12 µm/s
80 nM kinesin
v=24 µm/s
160 nM kinesin
v=6 µm/s
160 nMkinesin
v=12 µm/s
160 nM kinesin
v=24 µm/s
320 nM kinesin
v=6 µm/s
320 nM kinesin
v=12 µm/s
320 nM kinesin
v=24 µm/s
640 nM kinesin
v=6 µm/s
640 nM kinesin
v=12 µm/s
640 nM kinesin
v=24 µm/s
< AF (φ) >cycle
Cycle phase, φ
0
π/2
π
3π/2
2π
0.0
0.2
0.4
0.6
0.8

**Figure S8.** Alignment factor overlaid for all trials. $\langle AF(\varphi)\rangle_{cycle}$ is shown as a function $\varphi$ for each $(c_k, v)$ condition. In all cases, data corresponding to the closest ROI, $y$ = 10.5 μm, are shown. The thicker black line depicts the alignment factor averaged across all trials.

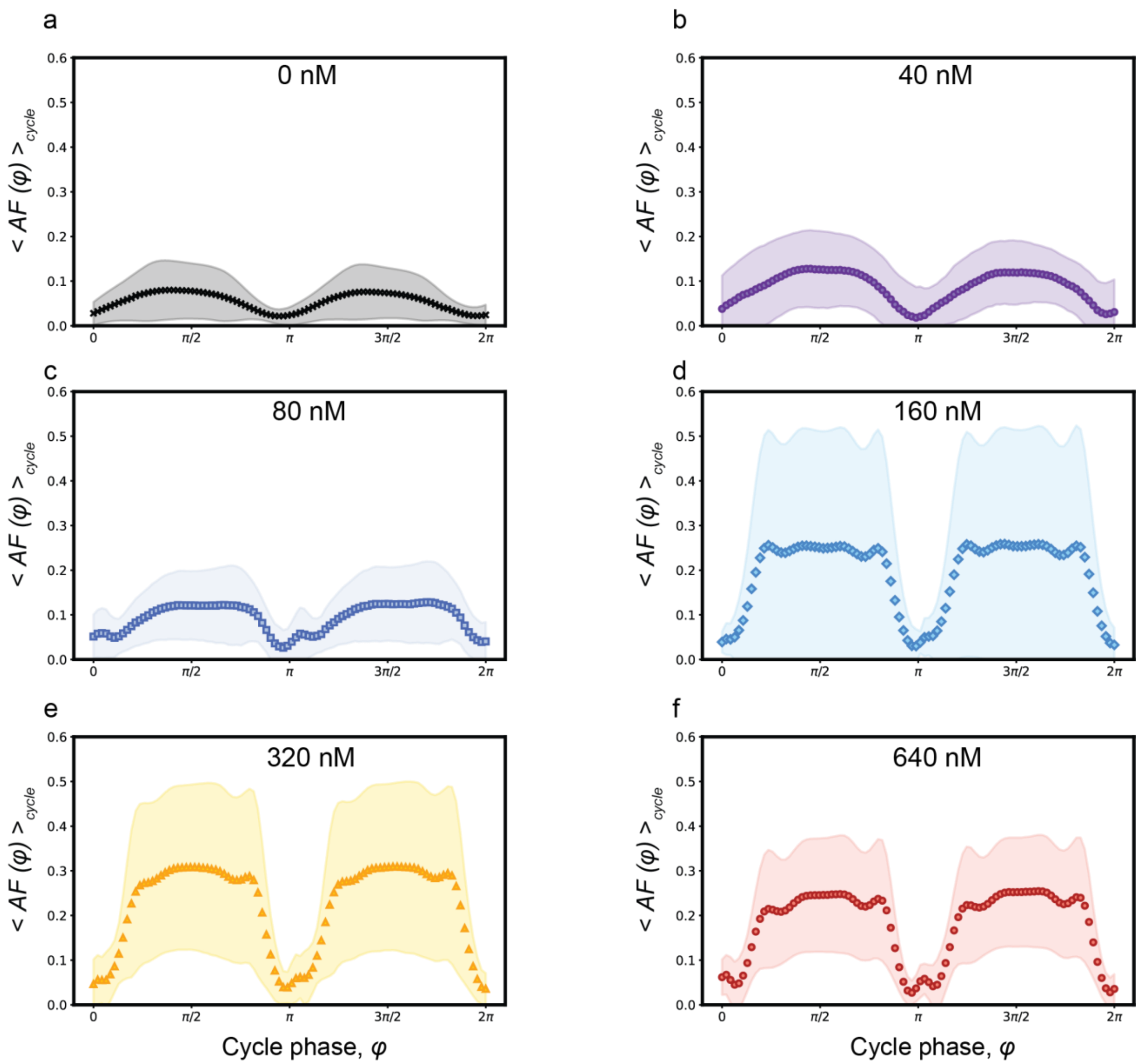

**Figure S9.** The alignment factors $\langle AF(\varphi)\rangle_{cycle}$ averaged across >20 trials are shown with solid symbols and standard deviation is shown as a shaded uncertainty band at (a) 0 nM, (b) 40 nM, (c) 80 nM, (d) 160 nM, (e) 320 nM, and (f) 640 nM kinesin concentration. Data shown are for $u$ = 12 μm/s and the closest ROI, $y$ = 10.5 μm. The largest uncertainty band occurs at $c_k$ = 160 nM, suggesting greater heterogeneity for this condition.

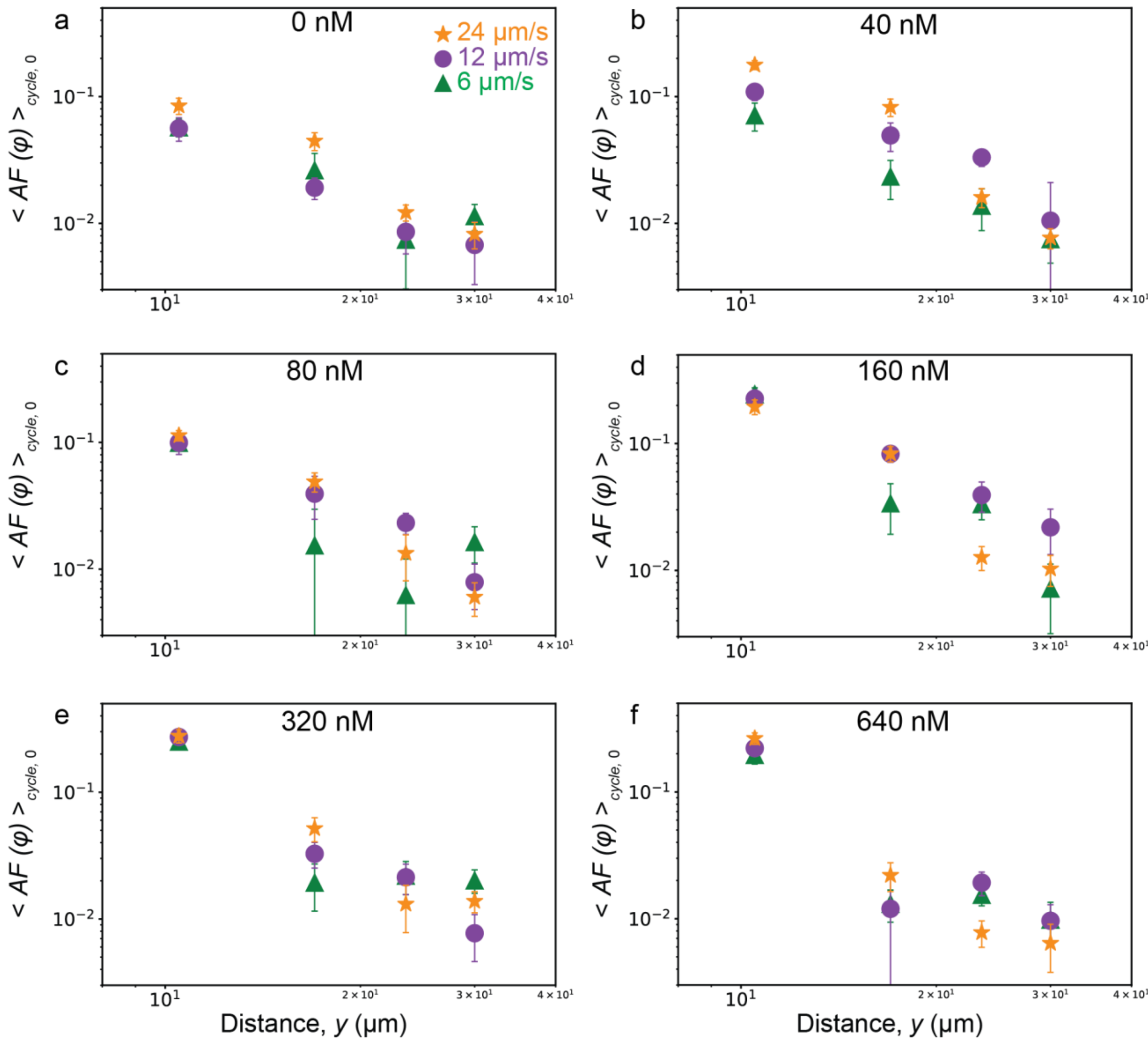

**Figure S10.** Effects of speed on the alignment factor obtained from time-resolved DDM. Green triangle symbols represent 6 μm/s, purple circles represent 12 μm/s, and orange stars represent 24 μm/s. Compared to the similar data of the time-averaged alignment factor, $\langle AF \rangle_t$, shown in Fig. S4, there are more discernable differences between the alignment factors for different speeds.